\documentclass[final,authoryear,5p,times,twocolumn]{elsarticle}
\providecommand{\tightlist}{%
  \setlength{\itemsep}{0pt}\setlength{\parskip}{0pt}}

\usepackage{amssymb}
\usepackage{hyperref}
\usepackage{booktabs}

\journal{Astronomy and Computing}

\begin{document}

\begin{frontmatter}



\title{Exploring the interpretability of deep neural networks used for gravitational lens finding with a sensitivity probe}


    \author[1,2]{C. Jacobs \corref{cor1}}
    \author[1,2]{K. Glazebrook}
    \author[3]{A. K. Qin}
    \author[4]{T. Collett}

    \address[1]{Centre for Astrophysics and Supercomputing, Swinburne
University of Technology, P.O. Box 218, Hawthorn, VIC 3122, Australia}
    \address[2]{ARC Centre of Excellence for All Sky Astrophysics in 3 Dimensions (ASTRO 3D),
Swinburne University of Technology, Hawthorn, VIC 3122, Australia}
    \address[3]{Faculty of Science, Engineering and Technology,
Swinburne University of Technology, P.O. Box 218, Hawthorn, VIC 3122,
Australia}
    \address[4]{Institute of Cosmology and Gravitation, University of
Portsmouth, Portsmouth PO1 3FX}
    \cortext[cor1]{colinjacobs@swin.edu.au; corresponding author}


\begin{abstract}
Artificial neural networks are finding increasing use in astronomy,
but understanding the limitations of these models can be difficult.
We utilize a statistical method, a \textit{sensitivity probe}, designed to complement established 
methods for interpreting neural network behavior by quantifying the sensitivity of a
model's performance to various properties of the inputs.
    We apply this method to neural networks
trained to classify images of galaxy-galaxy strong lenses in the Dark
Energy Survey. We find that the networks are highly sensitive to color,
the simulated PSF used in training, and occlusion of light from a lensed
source, but are insensitive to Einstein radius, and performance degrades
smoothly with source and lens magnitudes. From this we identify
weaknesses in the training sets used to constrain the networks,
particularly the over-sensitivity to PSF, and constrain the selection
function of the lens-finder as a function of galaxy photometric
magnitudes, with accuracy decreasing significantly where the \(g\)-band
magnitude of the lens source is greater than 21.5 and the \(r\)-band
magnitude of the lens is less than 19.
\end{abstract}


\end{frontmatter}


\hypertarget{sec:introduction}{%
\section{Introduction}\label{sec:introduction}}

Machine learning, the name we give to algorithms designed to learn from
data and make predictions without being explicitly programmed to do so,
is playing an ever-increasing role in modern astronomy \citep[for an
overview, see][]{flukeSurveyingReachMaturity2020}. As the volume of
``Big Data'' available for analysis increases rapidly
\citep{zhangAstronomyBigData2015, kremerBigUniverseBig2017}, more
efficient means of extracting scientifically relevant conclusions from
this data have become increasingly urgent. One machine learning
algorithm in particular---the artificial neural network
\citep[ANN;][]{rosenblattPerceptronaPerceivingRecognizing1957, fukushimaNeocognitronSelforganizingNeural1980}---has
proven to be successful in many applications within and without our
science. The power of ANNs lies in their ability to extract
task-oriented feature sets at different granularities, mapping from
input to output domains in a highly non-linear fashion and learning the
optimal functional form automatically from supplied data. They can be
scaled to almost arbitrary size and complexity. The now ubiquitous term
\emph{deep learning}
\citep{lecunDeepLearning2015, schmidhuberDeepLearningNeural2015} refers
to deep neural networks (DNNs), ANNs with many layers, and thus many
trainable parameters. Due to the advent of GPU computing and the
availability of large data sets, DNNs with up to billions of trainable
parameters are now routinely applied in domains such as computer vision
and natural language processing \citep{devlinBERTPretrainingDeep2019}.

One variant of DNN, the \emph{convolutional neural network}
\citep{LeCunBackpropagationAppliedHandwritten1989}, which is optimized
to exploit the relationships between neighbouring pixels in image data,
is extremely powerful in extracting meaning from images and has
revolutionized the field of computer vision
\citep{lecunConvolutionalNetworksApplications2010, krizhevskyImageNetClassificationDeep2012, voulodimosDeepLearningComputer2018}.

The application of deep learning to astronomy continues to accelerate.
Just a few examples include such disparate applications as cosmological
parameter estimation
\citep{ntampakaHybridDeepLearning2020, wangLikelihoodfreeCosmologicalConstraints2020};
gravitational wave identification
\citep{georgeDeepLearningRealtime2018}; galaxy morphology classification
\citep{dieleman_rotation-invariant_2015, zhuGalaxyMorphologyClassification2019, walmsleyGalaxyZooProbabilistic2020};
stellar classification \citep{sharmaApplicationConvolutionalNeural2019}
star-galaxy separation \citep{kimStargalaxyClassificationUsing2016}; and
photometric redshift estimation
\citep{hoyleMeasuringPhotometricRedshifts2016, eriksenPAUSurveyPhotometric2020}.
ANNs are even being used, albeit tentatively, to directly infer physical
laws, for instance by \citet{itenDiscoveringPhysicalConcepts2020} to
`rediscover' the heliocentric configuration of the solar system.

One area of astronomy where DNNs have had particular success is in the
discovery of strong gravitational lenses. The study of strongly lensed
galaxies is key in many areas of contemporary astrophysics, including
but not limited to cosmography
\citep{BonvinH0LiCOWNewCOSMOGRAIL2016, birrerH0LiCOWIXCosmographic2019, collettModelIndependentDeterminationH02019},
dark matter studies
\citep{Oguristellardarkmatter2014, LiConstraintsidentitydark2016, BirrerLensingsubstructurequantification2017},
the mass-assembly of lensing galaxies
\citep{sonnenfeldSL2SGalaxyscaleLens2013}, and the astrophysics of the
lensed sources themselves
\citep{jonesDustWindComposition2018, spilkerGalacticOutflowsHigh2019}.
However, at galaxy scale strong lenses are relatively rare, \(< 1\) in
1000 galaxies \citep{treu_strong_2010}, and are difficult to distinguish
from the galaxy population at large, displaying no large bias in color
or luminosity that enable them to be reliably singled out from
catalogs. Lenses can only be definitively identified using a
combination of color, morphology---such as Einstein rings/arcs,
multiple images of the background source---and spectroscopy. Automating
lens finding therefore requires either a significant investment of
expert human time, such as recruiting citizen scientists to examine
images
\citep{marshallSPACEWARPSCrowdsourcing2016, sonnenfeldSurveyGravitationallylensedObjects2020},
or a technique that can make full use of the morphological information
present in multi-band survey imaging. Previous attempts at automating
lens-finding in surveys involved the careful construction of algorithms
to detect rings and arcs
\citep{seidel_arcfinder:_2007, gavazzi_ringfinder:_2014, braultExtensiveLightProfile2015};
model sources as prospective lenses
\citep{marshall_automated_2009, ChanChitahStronggravitationallensHunter2015};
or a combination of these
\citep{sonnenfeldSurveyGravitationallylensedObjects2018}. These methods
resulted in some dozens of new discoveries. However, the extraordinary
success of of CNNs in computer vision generally
\citep{lecunConvolutionalNetworksApplications2010} makes them a logical
choice for the next generation of lens finders. CNNs have now been
successfully employed to discover new lenses in surveys such as the
Canada-France-Hawaii Legacy Survey
\citep{jacobsFindingStrongLenses2017}, Dark Energy Survey
\citep{jacobsExtendedCatalogGalaxy2019, jacobsFindingHighredshiftStrong2019},
Hyper Suprime-Cam Subaru Strategic Program
\citep{sonnenfeldSurveyGravitationallylensedObjects2018} and Kilo-Degree
Survey \citep{petrilloLinKSDiscoveringGalaxyscale2019}. A few thousand
new confirmed lenses or high-quality lens candidates have resulted from
these searches.

The scientific imperative for lens discovery is accelerating; rare
lenses, such as double or triple source plane configurations
\citep{collettTripleRolloverThird2020} can, even individually, provide
strong contraints on cosmological parameters, but are extremely
rare---of order one in \(10^6\)-\(10^9\) sources. Future pipelines, such
as the upcoming Legacy Survey for Space and Time
\citep[LSST;][]{ivezicLSSTScienceDrivers2019} will benefit from
real-time assessments of the presence of strong lensing. When a
transient candidate is identified, ideally a system will be in place to
instantly and reliably assess whether the host galaxy may be multiply
imaged by a foreground source and thus flagged for immediate follow-up.
In order for deep learning-based lens finding to drive this next wave of
scientific discovery, we will need to properly understand the
inefficiencies, biases and errors of our lens finders. For instance, the
selection function is not clear; is there a bias in the colors,
magnitudes, and other features of discovered lenses? How is the search
affected by the depth of the images or the seeing? 

Answering questions such as these is not straight forward. Despite their
successes in this and other fields, deep neural networks have a
significant drawback, namely their lack of interpretability. The mapping
of inputs to outputs that occurs in a deep neural network involves many
non-linear transformations parameterized by of order millions of
weights, making understanding the contribution of any feature or subset
of features of the input to the final determination very challenging. In
particular, the behaviour of a DNN when applied to an example that lies
outside the distribution used for training is undefined and
unpredictable. In computer vision, some attempts to interpret DNN
functioning have relied on \emph{salicency maps}, a family of techniques
that determine the most important (i.e.~salient) regions of the input in
making the final class determination. However, the utility of these
methods is limited, especially in a scientific/astronomical context,
since it is focused on producing a spatial saliency map without probing
other physical parameters that may be of interest to physicists. We
further discuss the difficulties with these techniques in
section~\ref{sec:interpreting} below.

In this work we examine two DNNs used to find strong gravitational
lenses in Dark Energy Survey
\citep[DES;][]{darkenergysurveycollaborationDarkEnergySurvey2016}
imaging and attempt to answer some of these questions. These networks
enabled the discovery over over 500 high-quality strong lens candidates 
\citep[]{jacobsExtendedCatalogGalaxy2019}, and 84 at redshifts $\sim0.8$ and 
above \citep[]{jacobsFindingHighredshiftStrong2019}. The networks were able
to produce samples for human inspection of considerable purity---up to one
in five of the most highly-scored galaxies were rated as probable or definite lenses.
However, the completeness of the search could only be estimated, as the
selection function of the DNNs could not be known.

Here we develop and employ a technique designed to probe the 
sensitivity of a deep neural network to various properties of the inputs. 
The method collects summaries of network performance across test sets,
and compares these performance summaries as a function of various
input parameters. We apply the method systematically to neural networks 
trained to find galaxy-galaxy strong lenses in survey
imaging, and test the sensitivity of the networks to color, 
PSF, noise, the occlusion of regions of the image, and the magnitudes of
lens and lensed source.

The paper is structured as follows. In section~\ref{sec:interpreting},
we provide some more detail on the challenges of interpreting neural
network outputs. In section~\ref{sec:methodology}, we summarize the
methodology used to probe the sensitivity of our trained networks to
various physical parameters. In section~\ref{sec:results} we detail the
results and discuss the implications of the insights learned for future
applications. Finally we offer concluding remarks in
section~\ref{sec:conclusion}.

\hypertarget{sec:interpreting}{%
\section{Interpreting neural networks}\label{sec:interpreting}}

Despite their success in many domains, not the least in astronomy, DNNs
suffer from serious interpretability problems. Describing what features 
a neural network is learning is not straight forward, and it is in 
general not possible to anticipate its failure modes, nor biases in its learned
features. In scientific applications quantifying these problems is of
increasing urgency.

Understanding the decisions of deep networks is a field of active research. 
A distinction is often made between \emph{interpretable} and \emph{explainable}
machine learning. An interpretable model is one which allows insight into
how it performs under certain circumstances, including potential failure modes.
An explainable model allows for a more detailed understanding of the model internals,
specifically the detailed reasons behind a particular decision. 
\citet[][]{gilpinExplainingExplanationsOverview2018} elaborate on the distinction
in more detail. Here we are focused on an aid to ANN interpretability.

\begin{figure*}
\hypertarget{fig:featuremaps_vgg}{%
\centering
\includegraphics[width=0.75\textwidth,keepaspectratio]{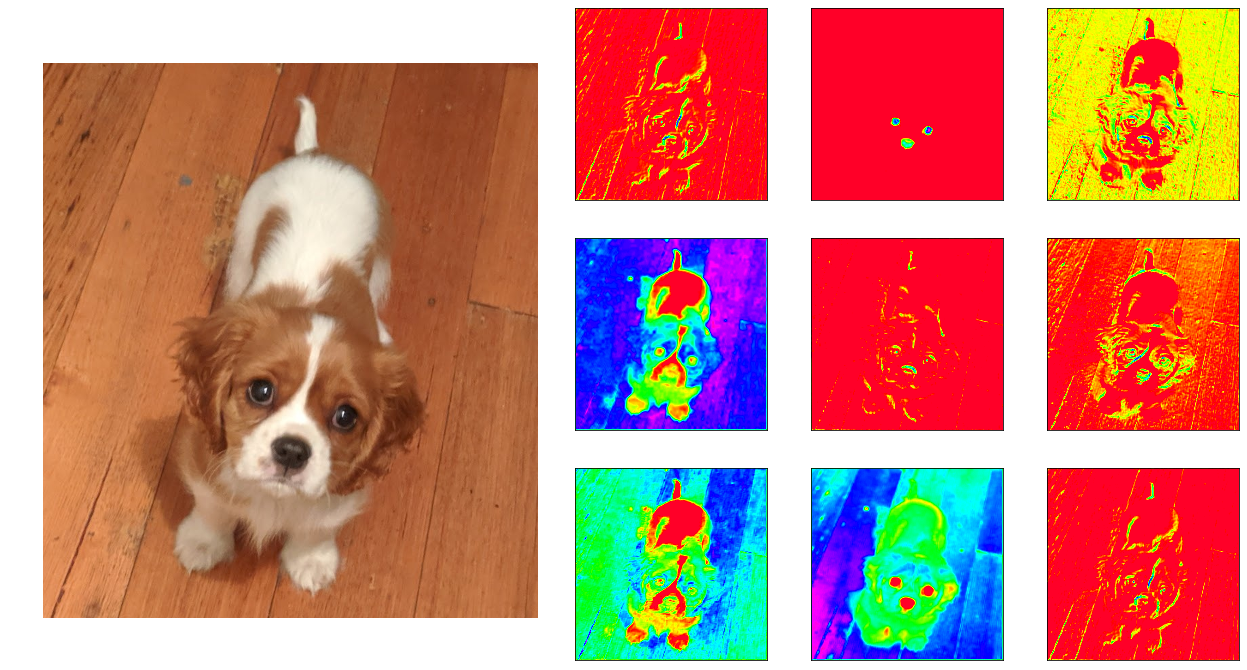}
\caption{Examples of feature maps from a convolutional neural network.
\emph{Left:} An image supplied to a network
\citep[VGG---][]{simonyanVeryDeepConvolutional2014} trained on images of
everyday objects. \textbf{Right}: Nine feature maps, the result of
convolving the input image with nine different convolutional kernels
from the first layer of the network, showing that the network has
learned to detect simple features at this level. Although these outputs
are interpretable, in that we can see how features such as edges are
detected by the network, it is difficult to draw a detailed
understanding of the network from such
examples.}\label{fig:featuremaps_vgg}
}
\end{figure*}
\begin{figure*}
\hypertarget{fig:featuremaps_lensing}{%
\centering
\includegraphics[width=0.95\textwidth,keepaspectratio]{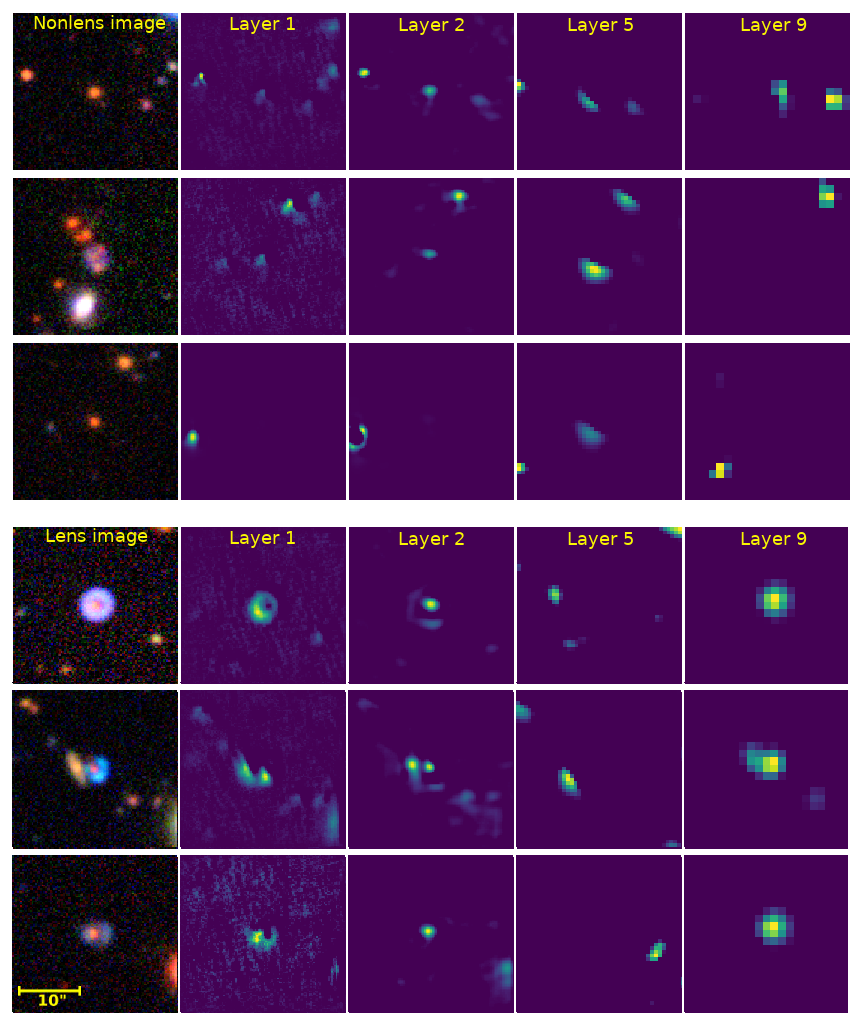}
\caption{Examples of feature maps from a convolutional neural networks
trained to detect strong gravitational lenses. \emph{Top:} Three
non-lens input images, convolved with convolutional kernels from
different layers of the network as indicated. \textbf{Bottom}: Three
simulated lenses and the resulting feature maps. Although the edges and
colors detected by the network are visible, a quantitative
understanding of network biases is difficult to extract from such
maps.}\label{fig:featuremaps_lensing}
}

\end{figure*}

Several approaches exist to interpret DNN behaviour. The most direct
approach involves visual inspection of the feature maps - the features
the network has been trained to recognise in a given input. At the
initial layer these features tend to be too simple to provide much
explanatory power; in the case of image data, for instance, vertical
edges or patches of color. At later layers the features are too
abstract to be interpretable visually, despite being rich in semantic
meaning. Figure~\ref{fig:featuremaps_vgg} shows an example from the
computer vision domain, for a neural network trained to recognise
everyday objects. At this early layer, we see the network has detected
features such as edges, but at a later layers spatial information has
been lost and the feature map is not interpretable.
Figure~\ref{fig:featuremaps_lensing} shows the same for networks trained
on simulated gravitational lenses, for both lens and non-lens images,
with example feature maps at several layers throughout the network.
These feature maps contain little or no quantitative information.

\begin{figure*}
\hypertarget{fig:saliencymaps}{%
\centering
\includegraphics[width=0.95\textwidth,keepaspectratio]{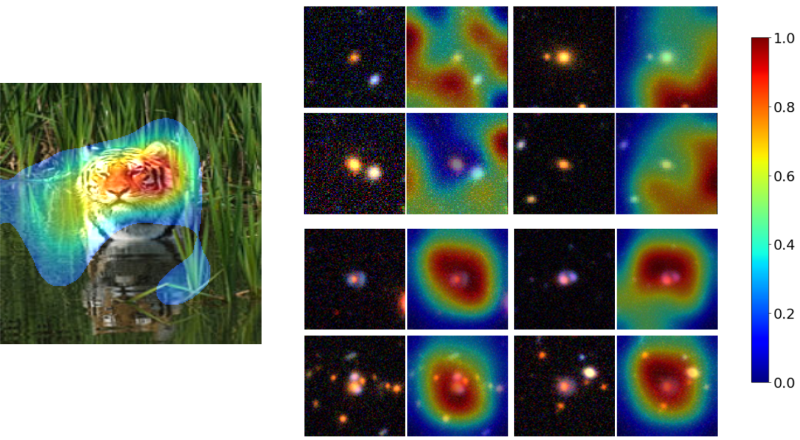}
\caption{Examples of saliency maps generated using the Grad-CAM
algorithm \citep{selvarajuGradCAMVisualExplanations2017}. \textbf{Left:}
An image of a tiger, passed to a network pre-trained on the ImageNet
dataset \citep{russakovskyImageNetLargeScale2015} with the salient
region depicted---the tiger's face. \textbf{Right:} Saliency maps for
simulated lensing and non-lensing galaxies. \emph{Top:} Four
non-lensing galaxies. The saliency is distributed throughout the image. \emph{Bottom:} Four lensing
galaxies; the salient region is the central source. 
The colors depict the most salient regions, from blue (lowest positive
saliency) to red (highest saliency).}\label{fig:saliencymaps}
}

\end{figure*}

With networks trained for computer vision, some of the most widely
adopted alternative approaches have focused on \emph{saliency
mapping}---determining and displaying those parts of the input most
crucial (salient) in determining a given output. Early approaches tried
calculating the signficance of individual inputs (pixels) to a class
score \citep[``Image-Specific Class Saliency
Visualisation''---][]{simonyanDeepConvolutionalNetworks2013}.
As the contribution of any individual pixel to the final score is
small, the outputs from this method were noisy and suffered from a lack
of ``dynamic range'' in explanatory power; at best, they indicate a
rather fuzzy, general area where there is a positive gradient with
regard to the correct class score \citep[see][]{simonyanDeepConvolutionalNetworks2013}.


Another variation on the theme is Gradient-weighted Class Activation Mapping
\citep[Grad-CAM;][]{selvarajuGradCAMVisualExplanations2017}, which examines the
feature maps at the last convolutional layer in a network, the point in
the network immediately before the output is flattened and all spatial
information is discarded. The algorithm weights these feature maps by
their contribution to the final score, then maps them back spatially to
the input image to produce a saliency map. Since CNNs usually discard
spatial information in favour of more feature maps at later layers, the
saliency map produced suffers from correspondingly lower resolution.
Other techniques develop these general ideas
\citep{smilkovSmoothGradRemovingNoise2017, zeilerVisualizingUnderstandingConvolutional2014, springenbergStrivingSimplicityAll2015, 
binderLayerwiseRelevancePropagation2016, kindermansLearningHowExplain2017},
but the central concept is the same; highlighting regions of interest in
an input image.

Are saliency maps useful in a scientific context? They are confined to
the exploration of spatial features, and typically lack granularity even
then. Figure~\ref{fig:saliencymaps} depicts, firstly, a Grad-CAM
saliency map generated on a neural network trained for visual
classification of photographs, applied to an image of a tiger \citep[as
per][]{selvarajuGradCAMVisualExplanations2017}. We can see that the
tiger's face is highly salient in determining the class depicted in the
image, an intuitive result. On the right, we show some saliency maps
applied to images of a strong gravitational lens, activated for the ``is
a lens'' class; it is equally intuitive that the central galaxy and
Einstein ring are salient in the determination, but does not provide
quantitative insight into the biases and weaknesses of the methodology.
The resolution of the saliency map is low, however it's not clear that
an improvement would allow significant new insights, as it is still
limited to the spatial dimension of the input only.

For the rest of this paper, we contrast saliency mapping techniques with
a \emph{sensitivity probe}, eschewing spatial information for patterns
derived from other known properties of the input, and demonstrate how
this approach may be more useful in quantifying the biases and strengths
of a neural network applied to the astrophysical problem of strong
gravitational lens detection.
The sensitivity probe is designed to be complementary to saliency maps,
particularly as an aid to the exploration of performance with regard to 
known physical properties.

\hypertarget{sec:methodology}{%
\section{Methodology}\label{sec:methodology}}

\hypertarget{data-sets-used}{%
\subsection{Data sets used}\label{data-sets-used}}

\hypertarget{dark-energy-survey}{%
\subsubsection{Dark Energy Survey}\label{dark-energy-survey}}

The networks analysed in this work were trained to detect strong
gravitational lenses in Dark Energy Survey
\citep[DES;][]{darkenergysurveycollaborationDarkEnergySurvey2016}
imaging. DES is an optical and near-infrared survey of 5000 square
degrees of sky in five bands (\emph{g, r, i, z} and \emph{y}). In
previous works
\citep[\citet{jacobsExtendedCatalogGalaxy2019}]{jacobsFindingHighredshiftStrong2019}
we searched the DES Year 3 coadd imaging
\citep{sevilla-noarbeDarkEnergySurvey2021}, and it is these images that
simulated lenses are designed to emulate. This imaging has a depth of
24.33 in \emph{g} (for a signal-to-noise of 10) and a pixel scale of
0.263 arcsec per pixel, with median PSFs of 1.12 arcsec FWHM in
\emph{g}, 0.96 in \emph{r} and 0.88 in \emph{i}.

\hypertarget{lens-models-and-test-set}{%
\subsubsection{Lens models and test
set}\label{lens-models-and-test-set}}

In the DES lens search we employed convolutional neural networks trained
using different training sets. The first, which we call ``Network A'',
was trained on simulated strong lenses generated using the
\textsc{LensPop} software \citep{collett_population_2015}, composed of
real images of large elliptical galaxies chosen from a catalog, combined
with a synthetic lensed source (henceforth, ``simulated lenses").
The negative examples comprise the
images of the elliptical galaxies only with no lensed source. The
second, ``Network B'', was trained on the same simulated lenses, 
with the negative examples comprising randomly chosen non-lens 
sources taken from the field, including elliptical and spiral galaxies,
mergers, and stars. The networks/training sets are summarized in Table~\ref{tbl:training_sets}.
For Network A, the presence of the lensed source is the 
key feature that indicates lensing, since the training set includes large
elliptical galaxies in both positive and negative examples. For network B,
the presence of a large elliptical (the most likely deflector in a strong
lensing system) is discriminative, but the network learns about spiral arms
and other potentially confusing features that are not features of lensing.
Here we contrast the sensitivity of these two training approaches to 
several input properties.

\hypertarget{tbl:training_sets}{}
\begin{table*}[]
    \centering
\caption{\label{tbl:training_sets} Summary of the training sets used to train
    the two networks (A and B), and the methodology used for positive and
    negative examples.}
\begin{tabular}{lll}
                        & \textbf{Lenses}                                                                                   & \textbf{Non-lenses}       \\
\textbf{Training set A} & \begin{tabular}[c]{@{}l@{}}Elliptical galaxies with simulated\\ lensed source added\end{tabular}  & Elliptical galaxies only       \\
\textbf{Training set B} & \begin{tabular}[c]{@{}l@{}}Elliptical galaxies with simulated \\ lensed source added\end{tabular} & Random sources from field
\end{tabular}
\end{table*}

We use a sensitivity probe on Networks ``A'' and ``B'' as
described above in section~\ref{sec:sensitivity}. The models each have
10 convolutional layers and two fully connected layers of 256 neurons
each. Each model has over 12 million trainable parameters. The training
sets used consisted of $\sim 150,000$ simulated lenses and a similar number
of non-lens galaxies. The output
distinguishes between the lens and non-lens classes, i.e.~produces a
probability that a tested image contains a gravitational lens. A further
test set, which also includes simulated elliptical galaxies, is used in
the test of PSF (section~\ref{sec:psf}).

Except as where noted, the test set we use consists of 5000 simulated
lenses, 5000 elliptical galaxies, 5000 real sources of all types, 
and 500 known lenses or high-quality lens candidates (human
verified, and with a high spectroscopic confirmation rate) from
\citet{jacobsExtendedCatalogGalaxy2019}. In all cases, we test images of
dimensions 100x100 pixels, corresponding to 26.3 arcseconds on a side,
in four bands (\emph{griz}). These images were not used during the training
process.

The purpose of the analysis is to better understand what the model
learned and identify weaknesses in the training set that could assist in
future searches. To that end, we apply the sensitivity probe,
examining the models' sensitivity to the following properties of the images (as
described in detail in \ref{sec:results}):

\begin{enumerate}
\def\labelenumi{\arabic{enumi}.}
\tightlist
\item
  \textbf{PSF:} We generate simulations with a PSF distribution that
  varies from that used in the simulations used to train the network,
    and also degrade the image with a Gaussian blur.
\item
  \textbf{Einstein radius:} The Einstein radius of the simulated lens.
\item
  \textbf{Galaxy magnitudes:} The \(g\)-band source magnitude and
  \(r\)-band lens magnitude.
\end{enumerate}

We also apply perturbations to the test set, to test the sensitivity to
the following parameters:

\begin{enumerate}
\def\labelenumi{\arabic{enumi}.}
\tightlist
\item
  \textbf{Noise}, by the addition of Gaussian noise;
\item
  \textbf{Color}, by artificially varying the colors of the images;
\item
  \textbf{Occlusion}, by zeroing out pixels in certain regions of the
  image
\end{enumerate}

\hypertarget{sec:sensitivity}{%
\subsection{Sensitivity probe}\label{sec:sensitivity}}

The technique used to examine the neural networks in this paper we refer
to as a \emph{sensitivity probe}.\footnote{Not to be confused with
  parameter sensitivity analysis, which tests the sensitivity to the
  weights of the network itself.} The key metric for the performance of
a neural network classification model is whether it puts the input in
the correct class. The output of a neural network trained for a
classification task is a vector of dimension \(n\), where \(n\) is the
number of possible classes, and the value \(s^i\), \(i \in [1...n]\) is
interpreted as a confidence that the input belongs to class \(i\). In
general the final layer includes a softmax activation such that
\(\sum_{i} s^i = 1\), so we can interpret \(s^i\) as a
(pseudo-)probability. In the optimal case, the output of the network
will be \(s^k = 1\) where \(k\) is the index of the correct (ground
truth) class, and \(s^i = 0\) where \(i \neq k\).

We use this correct-class score, \(s^k\), as the test for the sensitivity
probe. It is easily interpretable, as it represents the model's confidence
(in the range 0-1) that it has classified the input object in the correct
class category. Other metrics would be equally valid, such as the
categorical cross-entropy which is used as the loss metric to minimise
during the training process. This value is directly correlated with 
the model accuracy, but is less easily interpreted as it is 
unbounded as the correct-class score approaches zero.

We test how the correct-class score value varies across different test sets, or
as the parameters of a given test set are varied in some way. If this
value decreases for any given input or set of inputs, the performance of
the network can be said to degrade, and vice versa. If we can track this
degradation against some baseline, as the inputs are varied by some
parameter, then we can obtain a quantitive understanding of how
sensitive the performance of the network is to this parameter. 

In summary, the algorithm employed is as follows. We divide our test
set into subsets of approximately a few hundred objects, binned
by the property to be investigated. Then, the score in the range (0, 1) 
for the correct class, lens or non-lens, is predicted by the model for
each object. We calculate the mean score value $\mu$ per bin as well as the
standard deviation $\sigma$ in each bin. If the model performed perfectly
\(\mu\) and \(\sigma\) would be 1 and 0 respectively; in practice,
\(\mu\) is less than 1 and there is signifiant scatter in the predictions, 
representing both diversity in the test set and the inherent 
strength or weakness in the model when examining objects in a 
particular bin.
If the performance of a model degrades by bin, this will be 
reflected in a lower \(\mu\) (less confidence in
the correct class) and higher \(\sigma\) (more variation in score across the test
set). The sensitivity probe tests how these two quality metrics change across 
test sets in order to inspect how the score quality changes as a
function of the binning parameter. This is a purely empirical result, 
summarising the performance of the network; further statistical analysis,
such as performing Bayesian regression analysis of the relationship between 
score quality and input parameter against some prior, is possible. Here
we focus on the conclusions that can be drawn from the score quality data only.

The sensitivity probe also allows investigation using a ``perturber" function
instead of binning by some known parameter of the object. The perturber function
transforms each object in some way, for instance, by adding noise. This
allows one to test the sensitivity of the model to some property that is not
represented in catalog values to hand; in other words, we create a new distribution
of input objects that differs from the existing distribution in some way
and lets us see whether the model's score quality is sensitive to this shift.

The detailed algorithm for these two use cases, binning by parameter or
by peturbation, is described below.


For this analysis we use the \textsc{Sensie} software package
\citep{jacobsSensieProbingSensitivity2020} \footnote{https://github.com/coljac/sensie}.
\textsc{Sensie}, which is agnostic to the problem domain and
architecture of the trained model, automates the process of calculating
and plotting the accuracy of a neural network classifier controlled for
an input variable or perturber function.

\hypertarget{sensitivity-to-class-properties}{%
\subsubsection{Sensitivity to class
properties}\label{sensitivity-to-class-properties}}

The sensitivity probe determines the sensitivity of a (trained) model to some
scalar parameter, \(p\). This parameter may be some property of the
inputs; for example, the \(g\)-band luminosity of a galaxy, or even the
class index itself. In this case, we:

\begin{itemize}
\tightlist
\item
  Assemble a test set, \(T\), of objects with a known ground truth
  (correct class label).
\item
  Obtain the score given by the network for the correct (ground
  truth) class, \(s^k \in (0, 1)\) for each example in \(T\), by feeding
  it through the network.
\item
  Bin the results by values of \emph{p}; for each bin \(p\) calculate
  the mean score across the N examples in the bin:
  \[\langle s^{k}_{p}\rangle =\frac{1}{N} \sum_{i=0}^Ns^k_i\] as well as
  the corresponding standard deviation in \(s^k_p\).
\end{itemize}

\hypertarget{sensitivity-to-perturbation-of-the-input-data}{%
\subsubsection{Sensitivity to perturbation of the input
data}\label{sensitivity-to-perturbation-of-the-input-data}}

The sensitivity probe can test the sensitivity to a \emph{perturbation} of the
data, an arbitrary transformation applied to each example in T,
parameterized by a scalar \(p\), such that \(T' = f(T, p)\) for a
supplied perturbation function \(f\). For example, we may wish to
consider the sensitivity of the network to the rotation of input images;
this could be parameterized by the rotation angle. In this case, the
algorithm is as follows:

\begin{itemize}
\tightlist
\item
  Choose set of discrete values of \(P\) for testing over some range,
  \textbf{P}.
\item
  For each value in \textbf{P}, \(p\), obtain \(T' = f(T, p)\).
\item
  Calculate the score given by the network for the correct (ground
  truth) class \(k\), \(s^k_i \in (0, 1)\) for each example \(i\) in
  \(T'\).
\item
  Bin the results for each value of \(p\) and calculate the mean score
  \(\langle s^k_p \rangle\) by bin, as well as the corresponding
  standard deviation of \(s^k_p\).
\end{itemize}

In each case we have a set of discrete values or bins of some parameter
\(p\), and a measure of the performance of the network corresponding to
data represented by this value, \(\mu_p\) and \(\sigma_p\). We may then
investigate how the performance of the network varies as a function of
\(p\). This can be done through visual inspection of the results,
however we can also perform linear
regression on \(p\) versus \(\mu_p\) to test whether the effect is
significant---e.g. whether a zero gradient is consistent with the data.
For an example, see
section~\ref{sec:einstein}.

\hypertarget{sec:results}{%
\section{Results and discussion}\label{sec:results}}

\hypertarget{colour-1}{%
\subsection{Color}\label{colour-1}}

\subsubsection{Test performed}

We vary the colors of the images by applying a random ``jitter'' or
scaling factor independently to the three bands (\emph{g}, \emph{r}, and
\emph{i}) consumed by the network. The data in each band is scaled by a
value drawn from a normal distribution with a mean of 1 and a standard
deviation \(\sigma\) between 0 and 1. For each value of \(\sigma\), in
increments of 0.05, we draw a scale factor for each band in image in the
test set and apply the scaling (we multiply all pixels in the band by
this factor), before testing the image to determine network accuracy. We
thus test the sensitivity of the model to the amount of variation in the
relative scaling of the three bands parameterized by \(\sigma\),
i.e.~the fidelity of color of the image to the original simulation or
observation.

The addition of this color jitter does not represent an astrophysically
meaningful variation in the colours of the galaxies, which would be 
best accomplished by testing with a wide array of templates, stellar
populations and/or star formation histories. The purpose of this
test is simply to see whether the machine learning method is highly 
sensitive to the distribution of colors provided during training.
A high sensitivity to the `correct' colours would indicate that
the model would benefit from seeing a wider, and physically motivated,
distribution of galaxy colors in the training step.

\subsubsection{Results}

\begin{figure*}
\hypertarget{fig:jitter}{%
\centering
\includegraphics[width=0.95\textwidth,keepaspectratio]{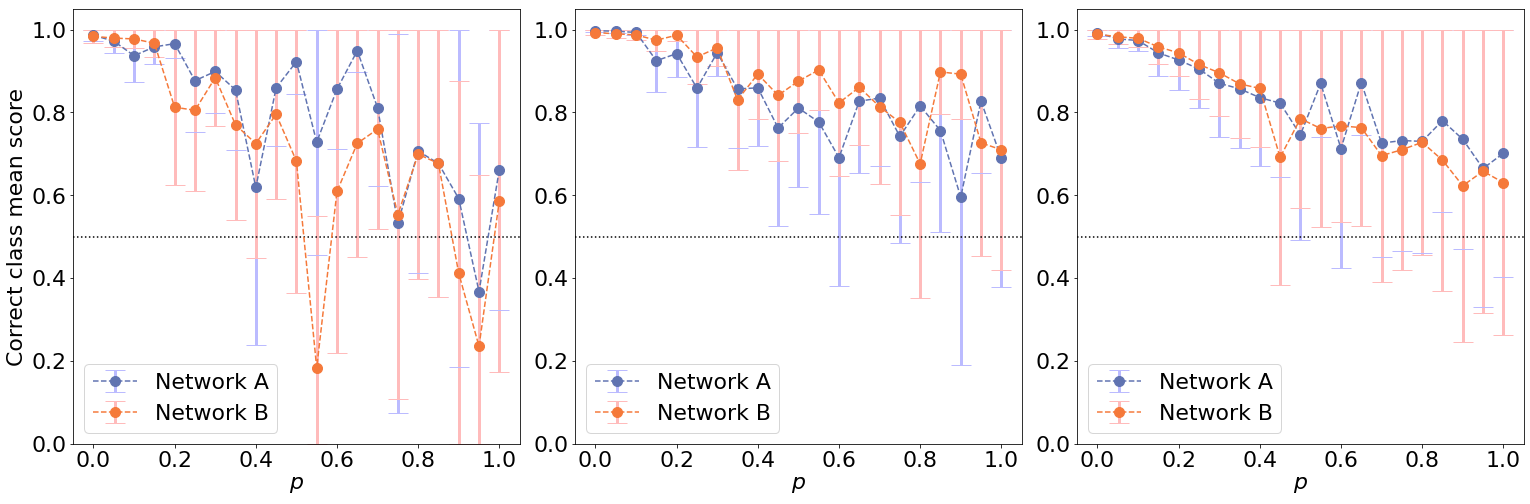}
\caption{The effects of adding color `jitter' to lens images,
i.e.~arbitrarily changing the scaling between bands in an image. Here
the magnitude of the effect is parameterized by \emph{p}, where \emph{p}
corresponds to the mean size of the effect, from 0 to 100\% random
variation. \textbf{Left}: Effect on the two networks when applied to
simulated strong lenses. \textbf{Middle}: Effect on non-lens elliptical
galaxes (Network A) and other non-lens real galaxies (Network B).
\textbf{Right}: Combined effect on a test set containing 50\% lenses and
50\% non-lens images.}\label{fig:jitter}
}

\end{figure*}

The network was sensitive to the fidelity to the original colors of the
input images. Figure~\ref{fig:jitter} (left) shows the sensitivity of
the two networks to color, when tested on test sets consisting either of
simulated lenses, non-lensing galaxies, or a mixture of both.
The plots depict the mean score given by the network for the correct class
(lens/non-lens), with error bars showing one standard deviation. On the
left, we see the effect on the scores of simulated lenses for
network A, trained on simulated lenses and elliptical galaxies, 
and network B, trained on simulated lenses and a diverse selection of real
galaxies. The performance is similar, although network B
appears slightly more robust at small values of the jitter factor. In
both cases, the networks remain more than 90\% confident in the lens
images until the jitter factor exceeds 20\%, at which point the accuracy
decreases quickly. At a jitter factor higher than .2 in the case of
network B and .4 in network A, the threshold of 0.5 is less than one
standard deviation away from the mean score---the network is very
unreliable at this point.

On non-lensing galaxies, the performance 
degrades similarly. The middle panel of Figure~\ref{fig:jitter} shows
the performance of the network A on non-lens elliptical galaxies and
network B on other real field galaxies. A score of 1 in this case
indicates certainty that the object is not a lens. The degradation in
performance is similar to the lenses. Thus, when it comes to color
information, deviating from the simulated colors used in the training
set, or from the colors of real field galaxies, leads to confusion in
the networks - both lenses and non-lenses become increasingly uncertain.

Figure~\ref{fig:jitter} (right) shows the accuracy of the networks on a
test set consisting of 50\% simulated lenses and 50\% non-lensing
galaxies, i.e.~combining the effects of the two separate test sets. The
performance degrades linearly until the jitter reaches a factor of
approximately 0.5, at which point the accuracy of the network has
decreased from $\sim 100\%$ to $\sim 75\%$.

From the above we conclude that the network has learned that the colors
of the objects in the image are determinative of lensing. Performance
may be improved by introducing a wider range of colors into the
simulations used for training, in case the network becomes overly
sensitive to the choices made here. For instance, the colors of lensed sources
in the simulations were drawn from star-forming galaxies in the COSMOS catalog
\citep{ilbertCosmosPhotometricRedshifts2009}, and when using simulated 
elliptical galaxies in some experiments, we relied on a single
template for a 10-gigayear-old passive galaxy.
Re-training with a wider family of galaxies
(both lenses and source) may reduce the effect of this sensitivity to
color and enable the discovery of more lenses with atypical colors,
for instance red-red lenses.

\begin{figure*}
\hypertarget{fig:blur}{%
\centering
\includegraphics[width=0.95\textwidth,keepaspectratio]{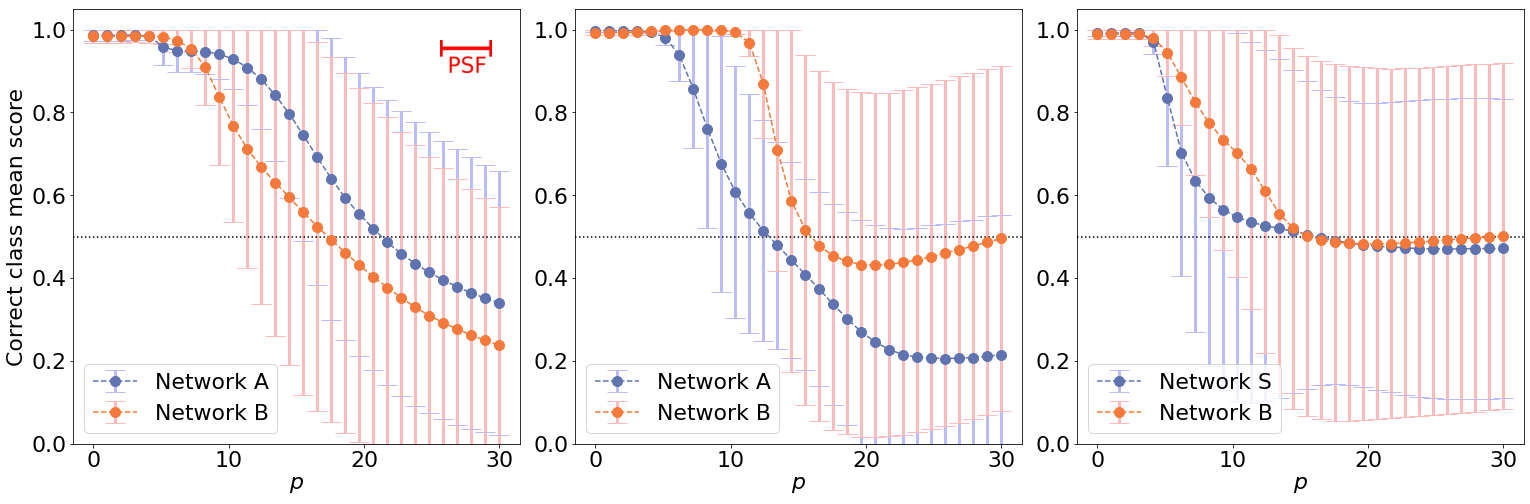}
\caption{The effects of blurring test images by convolving with a
Gaussian kernel. Here the magnitude of the effect is parameterized by
\emph{p}, where \emph{p} corresponds to the RMS of the kernel in pixels.
The mean PSF size (1.0 arcsec) is indicated on the plot for reference.
\textbf{Left}: Effect on the two networks when applied to simulated
strong lenses. \textbf{Middle}: Effect on non-lens images for Networks A and B.  
\textbf{Right}: Combined effect on a test set containing 50\% lenses and 50\% non-lens
images.}\label{fig:blur}
}

\end{figure*}

\hypertarget{sec:psf}{%
\subsection{PSF}\label{sec:psf}}

\subsubsection{Test performed}

Our simulated strong lenses take an image of a real elliptical galaxy,
estimate a realistic lensing mass, and simulate an image of a background
source at higher redshift. This image is convolved with a PSF drawn from
a distribution designed to match the survey properties and then added to
the real image of the simulated galaxy.
The simulations on which our model was trained use a PSF modeled as a
Gaussian with a FWHM drawn from a distribution with means of 1.27, 1.08
and 0.98 arcseconds in \emph{gri} respectively, consistent with the DES
Science Verification imaging. To test the sensitivity to this parameter,
that is, to determine if the trained model is more or less likely to
recognize a gravitational lens when the PSF of the image matches the
initial training set, we create further simulations, with synthetic lens
and source galaxies, using PSFs drawn from a distribution with a mean of
between \(\sim 0.7\) and 3.3 arcsec, i.e. $\sim 0.6$ to 3 times that used in
the training set.

For this test, we created two test sets composed of new simulations with
this property, one containing simulations of the sort used to train the
networks, with real galaxies as lenses and simulated sources 
(section~\ref{sec:sensitivity}); the other using both simulated sources
and simulated elliptical galaxies in order to provide an additional
test. This was done to create realistic images that have slightly
different properties to the training set, as would occur when evaluating
real galaxies with the model that have parameters not fully captured in
training. The simulations were once again generated with
\textsc{LensPop}.

Since the PSF only effects the simulated aspects of the images---the lensed source,
and in the case of the test set described above, also a synthetic elliptical lensing 
galaxy---we also probe the effect of convolving the image with a 
2D Gaussian kernel with a standard deviation, degrading the entire image including all foreground
and background sources. We employ a Gaussian with standard deviation,
\(\sigma\), between 0 and 3 pixels (0-0.8 arcsec). We test how the performance of the network
degrades as a function of this parameter \(\sigma\). With this test, we
hope to get an indication of the scale of the significant features used
to determine whether a source is a strong lens or not.

\begin{figure*}
\hypertarget{fig:perturbed}{%
\centering
\includegraphics[width=0.95\textwidth,keepaspectratio]{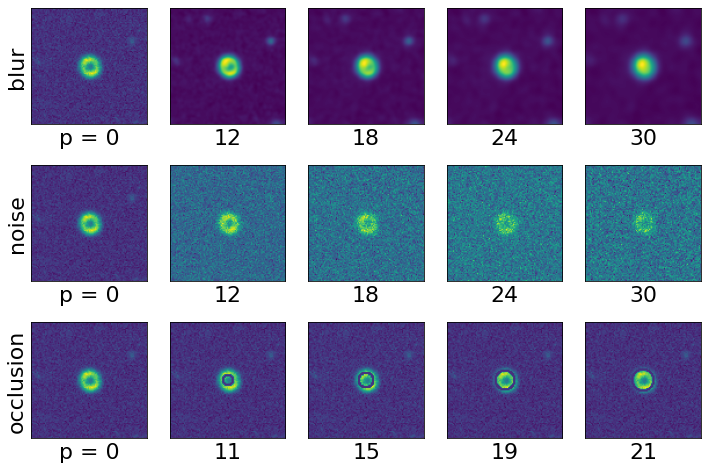}
\caption{An example simulated lens from the test set before and
after perturbation with one of the perturber functions. In order from
top: Convolution with 2D-Gaussian (blur), with Gaussian width 0 to 3
pixels; addition of Gaussian noise, degrading S/N from 14 to
$\sim 2$; occlusion of an annulus 5 pixels wide with radius
shown (pixels). Images are 100 pixels across ($\sim 26$ arcsec
in DES).}\label{fig:perturbed}
}

\end{figure*}

Figure~\ref{fig:perturbed} depicts an example test set image that has
been perturbed as described above with the blur, noise and occlusion
perturbers.

\subsubsection{Results}

As expected, we find that the Gaussian
blurring of the images degrades performance of the network. However, the
response to the blurring is not entirely consistent across the networks.
Figure~\ref{fig:blur} (left) shows the response of the networks when
tested on simulated strong lenses with blur applied; this can be
contrasted with the middle panel, which shows the response of the
networks to simulated and real non-lenses. For both networks, increasing
blur decreases the certainty of lensing, and this effect continues
beyond the 50\% threshold (the point at which the network would be
guessing randomly between the two classes); i.e.~the blurrier the image,
the lower the probability of lensing assigned to the networks. The rate
of this decline is different, however both networks show robust
performance until the width of the kernel reaches $\sim 5$
pixels, corresponding to an angular scale of 1.3 arcsec. This,
presumably not coincidentally, is the approximate Einstein radius at
which lensing becomes detectable in DES imaging.

In the case of the non-lenses in Figure~\ref{fig:blur} (middle panel),
network B degrades to approximately 50\% and remains there, however
network A declines below the 50\% threshold. In either case, the
performance on a balanced test set containing a 50/50 split (rightmost
panel of Figure~\ref{fig:blur}) degrades smoothly to the expected 50\%
threshold.

Although the results are not surprising---applying a blur removes
information from the image, and so the accuracy must necessarily
decrease--one conclusion we can draw from this is that the networks have
incorporated the features of the two training sets in different ways.
Network A, trained only on simple simulated non-lenses as negative
examples, thinks that non-lenses are more likely to be lenses the worse
the resolution becomes. From this we can infer that it is quite
sensitive to the properties of the simulated early-type galaxies that we
used, such as the surface brightness profile, as well as the resolution
of the simulated DES images. Once it is shown examples that deviate from
this---even if no lensed source is introduced---it becomes increasingly
confident the example is a lens. By contrast, its confidence in a
blurred lens degrades more slowly. Network B, perhaps more intuitively,
loses confidence in lenses as they become more distorted, but for non
lenses it dips only slightly below the 0.5 ``random guess'' threshold.

\begin{figure*}
\hypertarget{fig:psf}{%
\centering
\includegraphics[width=0.95\textwidth,keepaspectratio]{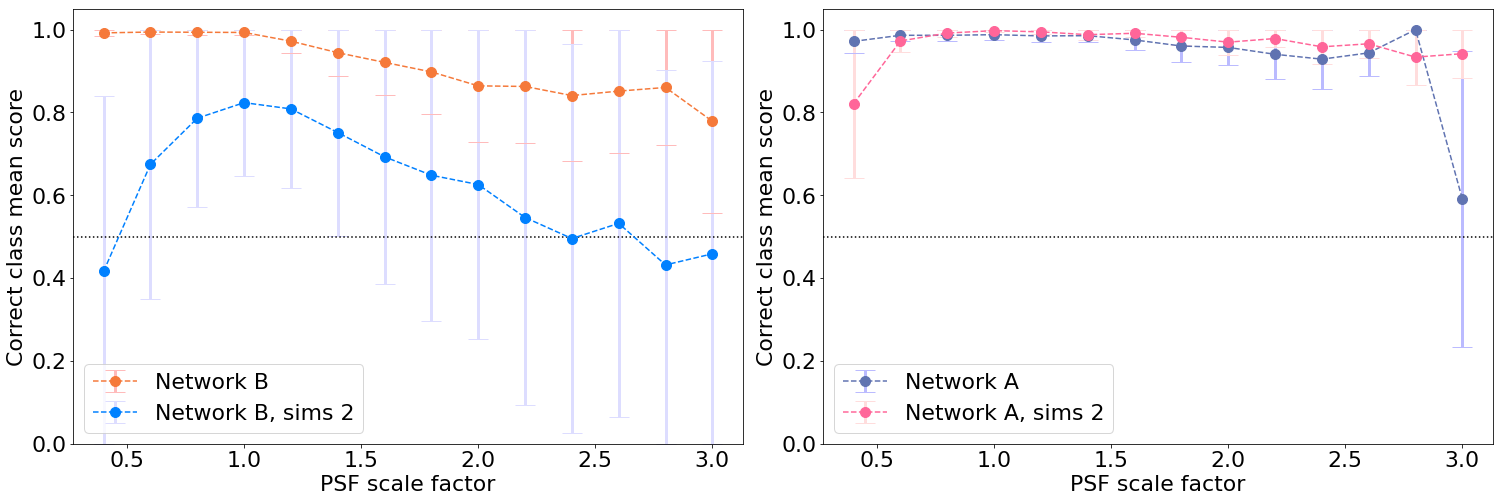}
\caption{The effects of the PSF in simulated images on the network
accuracy, Here the magnitude of the effect is parameterized by \emph{p},
where \emph{p} corresponds to the ratio of the test set PSF mean to the
mean PSF used to train the networks originally. \textbf{Left}: Effect on
the Network B, trained on simulated strong lenses and real field
galaxies. \textbf{Right}: Effect on network A, trained on simulated
lenses and elliptical galaxies. In both cases, the performance of the
networks decreases when the PSF is \emph{better} than the fiducial value
used for training.}\label{fig:psf}
}

\end{figure*}

With regards to the PSF, the results are more instructive.
Figure~\ref{fig:psf} shows the response of the networks to simulated
strong lenses with different PSFs and two different simulation
methodologies. The first type of simulations are the same as those used
to train the networks (simulated lensed sources and catalog elliptical
galaxies, see Table~\ref{tbl:training_sets}); the second, 
labelled `sim2' in the figure, use simulated
elliptical galaxies. The x-axis represents a scale factor relative to
the fiducial PSF used to create the simulations on which the networks
were trained, a mean of approximately 1 arcsec. As the PSF gets wider,
so the performance of the network degrades for both test sets, an
expected result, although the effect is much larger for Network B.
However, in the cases of both networks, a PSF better than the training
set value also leads to decreased performance. One interpretation of
this result is that the network is `smart' enough to eschew lenses with
unphysically sharp features. However, this also indicates a strong
sensitivity to the simulated values. In the case of the lens search
conducted in \citet{jacobsExtendedCatalogGalaxy2019}, the simulations
used a simulated PSF designed to match that reported in the DES Science
Verification data ($\sim 1$ arcsec), but the search was
conducted on Year 3 coadd images, which had a better PSF
($\sim 0.8$ arcsec). From this experiment we conclude that
several lenses may well have been missed due to this effect. This
suggests a serious weakness in the training set that could be remedied
firstly by better matching the PSF to the target imaging, but also by
introducing a wider distribution of PSFs into the training set, forcing
the network to adapt to a wider range of conditions.

\hypertarget{sec:einstein}{%
\subsection{Einstein Radius}\label{sec:einstein}}

\subsubsection{Test performed}

The Einstein radius, which is a function of the geometry of the lensing
system and the lens mass, is calculated when the lenses are simulated.
This data is available for simulated lenses only; for genuine
lenses and lens candidates the Einstein radius is not precisely known (nor easily
calculated) without high-resolution imaging. Although we can only test
simulations this way, we can still obtain insights into how well the
lensfinder works for less obvious lenses (smaller Einstein radius), and
for those that are less represented in the training set.

We create a test set of 10000 simulated lenses with an Einstein radius
\(1.0 < E_r < 2.7\), a typical range for likely, detectable lenses in
Dark Energy Survey DECam imaging. We bin the sims by Einstein radius,
using 30 bins of 0.052 arcsec wide (approximately 300 per bin) and
assign each bin's midpoint to the simulations in that bin. These bin
values are passed to the sensitivity probe.

\subsubsection{Results}

\begin{figure*}
\hypertarget{fig:einstein}{%
\centering
\includegraphics[width=0.95\textwidth,keepaspectratio]{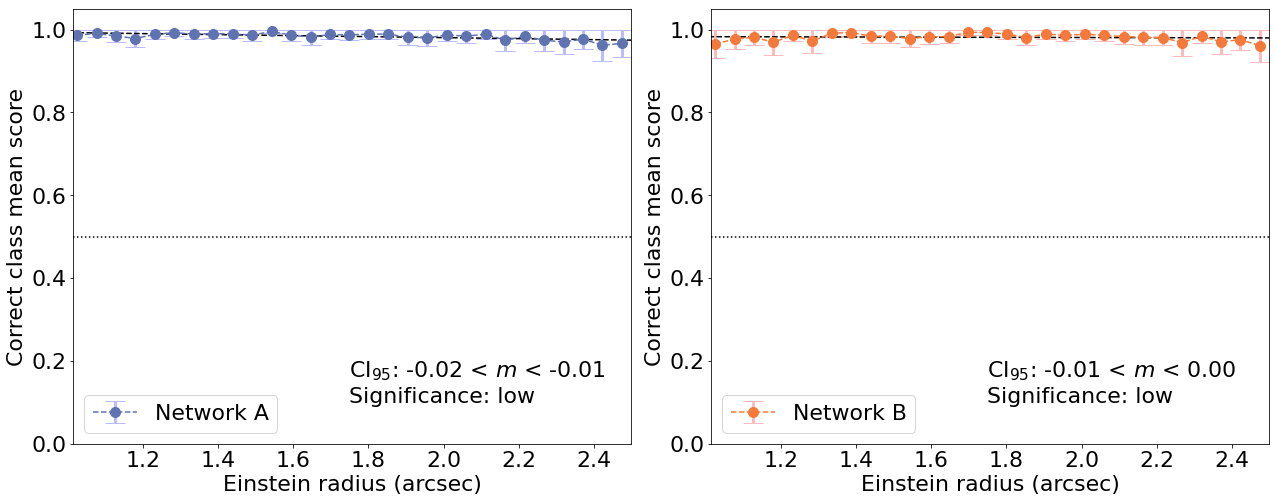}
\caption{The effects of the simulated lens Einstein radius on network
accuracy. Here we compare the mean score given by the networks to
simulated lenses binned up by Einstein radius, with bins 0.052 arcsec
wide/ \textbf{Left}: Effect on the Network A, trained on simulated
lenses and elliptical galaxies. \textbf{Right}: Effect on network B,
trained on simulated lenses and real field galaxies. Both networks
display a strong insensitivity to this parameter. The 95\% credible
interval for a the gradient of a linear fit to the data is shown; it is
consistent with zero or nearly zero in both cases.}\label{fig:einstein}
}

\end{figure*}

For this test we pass the networks 10,723 simulated lenses with known
Einstein radii \(R_E\). Figure~\ref{fig:einstein} depicts the
sensitivity of the networks to Einstein radius. On the left, network A
and on the right, network B. In the first case, there appears to be a
slight decrease in lens certainty as \(R_E\) increases; in the second, a
slight increase up to 1.75 arcsec, then a decrease. The magnitude of the
effect is low; in the case of network B, assuming the standard deviation
as errors in the outputs, the data is consistent with a
zero gradient inside a 95\% credible interval. In the case of network
S, the gradient, \(\partial \langle s^k \rangle/\partial R_E\) is
significant but small, with mean accuracy increasing at 1.2\% per
arsecond across the tested interval.

Understanding the different response between the two networks may
require further experimentation. Network A was only trained with images
of elliptical galaxies, and so barring a few coincidental aligments in
the training set---the simulated galaxies were placed in DES tiles so
that interlopers, stars and artifacts were present---the presence of any
arc is likely to be indicative of lensing. In the case of Network B,
this is not the case, as the training set contained many thousands of
spiral galaxies. Smaller Einstein radii were over-represented in the
training set, whereas spiral arms could reach arbitrary size in the DES
imaging. Thus, at larger radii the network becomes less certain, as the
risk of confusion with a spiral arm or tidal tail is larger. In any
case, this test does not reveal a deficiency or over-sensitivity in the
training set for either network.

We note results from real (human inspected) lens candidates are not
included here as the Einstein radius is not known and is difficult to
constrain with ground-based imaging.

\hypertarget{lens-and-source-magnitudes}{%
\subsection{Lens and source
magnitudes}\label{lens-and-source-magnitudes}}

\subsubsection{Test performed}

For simulated lenses, we know the (observed) magnitudes of the lens and the
source. Here, we test the sensitivity of the
lens-finder to the \(r\)-band magnitude of the lensing galaxy and the
\(g\)-band magnitude of the lensed source. To test the sensitivity to
these parameters, we collect 4000 simulated lenses where
\(19 < r < 21\) and 4000 simulated lenses where 
\(20 < g_{\textrm{lensed}} < 22\). We create 20 bins 0.1 magnitudes wide
across each magnitude domain, with each containing \(\sim 200\) sources
respectively. The midpoint of the magnitude bin is passed to
the sensitivity probe as the parameter to test.

\subsubsection{Results}

\begin{figure*}
\hypertarget{fig:mags}{%
\centering
\includegraphics[width=0.95\textwidth,keepaspectratio]{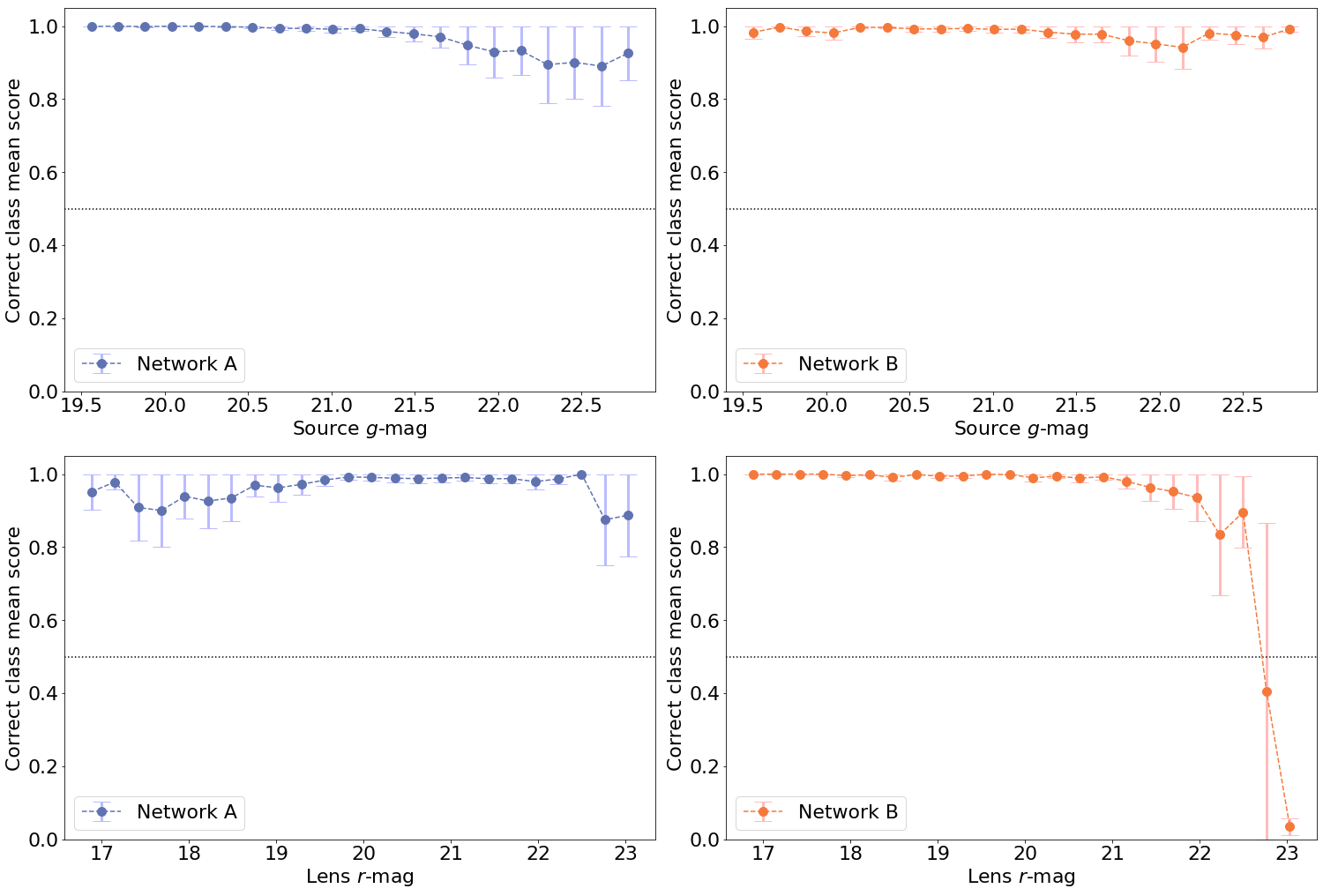}
\caption{The accuracy of the networks in detecting simulated lenses, as
a function of galaxy magnitude. \textbf{Top left}: Network A, trained on
simulated lenses and elliptical galaxies, as a function of the
integrated \(g\)-band magnitude of the lensed source. \textbf{Top
right}: Network B, trained on simulated lenses and real field galaxies.
\textbf{Bottom left}: Network A, accuracy as a function of the lens \(r\)-band magnitude.
\textbf{Bottom right}: Network B, Accuracy as a function of the lens \(r\)-band
magnitude.  }
    \label{fig:mags}
}

\end{figure*}

By examining the sensitivity to the magnitudes of source and lens, we
hope to gain a better understanding of the selection function of the
lens-finder. Figure~\ref{fig:mags} depicts the response of the two
networks to simulated lenses as a function of lens \(r\)-band and source
\(g\)-band magnitude. The magnitude ranges explored reflect the
parameters of detectable lenses used to train the networks.

The response of the two networks are somewhat dissimilar; network B is
less sensitive across most of the range. Network A degrades in
performance as the source gets fainter, with significant effects from a
\(g\)-band magnitude of 21.5. Network B becomes less accurate at a
similar range, although the trend is less clear. With the lens
\(r\)-mag, Network A shows a significant degradation in performance
where the lens is very bright, \(r < 19\); this is not evident for
network B. Network A also showed a degradation in performance for very
faint lenses, where \(r > 22.5\). Network B displays no sensitivity to
bright lenses but fainter than 21 in \(r\) performance degrades quickly.

In the \citet{jacobsExtendedCatalogGalaxy2019} search, we conducted the
search by examining candidates above a certain score threshold (for
instance, 0.99---very certain candidates) and then lowering the
threshold in increments until the rate of discovery makes further
examinations no longer worthwhile. From the sensitivity probe, we can
map these thresholds to a selection function: At threshold .9, network A
is unlikely to find many lenses with a \(g\)-band magnitude greater than
22. The scatter in the scores makes it difficult to establish a firm
cutoff, but for a given score threshold we can establish a point where
$\sim 50$\% of lenses are unlikely to be found for any given
magnitude value.

\begin{figure*}
\hypertarget{fig:mags2}{%
\centering
\includegraphics[width=0.95\textwidth,keepaspectratio]{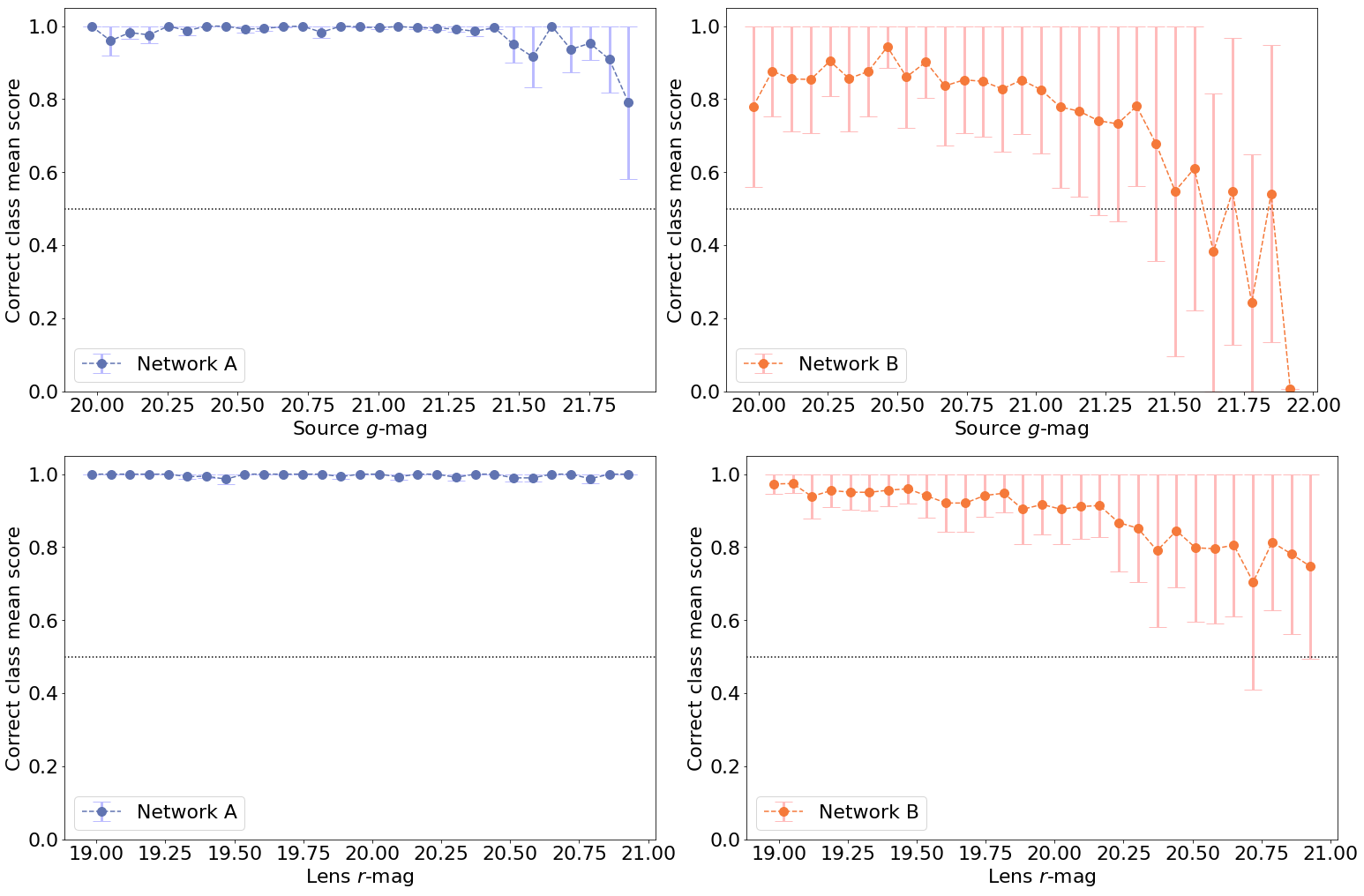}
\caption{The accuracy of the networks in detecting simulated lenses, as
a function of galaxy magnitude, for simulations containing synthetic
elliptical galaxies. \textbf{Top left}: Network A, trained on simulated
strong lenses, as a function of the integrated \(g\)-band magnitude of
the lensed source. \textbf{Top right}: Network B, trained on simulated
lenses and real field galaxies. \textbf{Bottom}: Accuracy as a function
of the lens \(r\)-band magnitude. \textbf{Left}: Network A,
\textbf{Right}: Network B.}\label{fig:mags2}
}

\end{figure*}

This experiment was repeated using simulations developed with a
different methodology, as per section~\ref{sec:psf}. Results of this
test are presented in Figure~\ref{fig:mags2}. Although these simulations
are less realistic than those described above, this may serve to
highlight further weaknesses of the network by showing them examples
dissimilar to those used in training, and thus less susceptible to
over-fitting to features of the simulations. This analysis provides some
evidence that of the limitations of a network trained only on one type
of galaxy (bright ellipticals). The high accuracy for network A even for
faint objects indicates that without having to account for the variety
of morphologies and colors found in galaxies in general, the network
can likely make some assumptions that will not survive contact with
diverse (and lower signal-to-noise) sources from the field. 
In
\citet{jacobsExtendedCatalogGalaxy2019}, we used network B specifically
to account for weaknesses in the simulations: a network trained with the
same elliptical galaxies in both the positive and negative training sets
could not learn that some feature of these galaxies alone was
discriminative. 
However, the use of network B, which was exposed to the
full range of non-lens galaxy morphologies in training, was essential to
balance the unrealistic simplicity of the all-simulation training set.

\hypertarget{occlusion-1}{%
\subsection{Occlusion}\label{occlusion-1}}

\subsubsection{Test performed}

Occlusion---that is, masking out parts of an image to determine the
effect on the score---has been used as a simple form of saliency mapping
(see section~\ref{sec:interpreting}). In theory, one could zero out each
pixel of an image one by one and calculate the effect on the network's
score, and thereby create a map of \(\partial \hat{y}/\partial p_{ij}\)
over the image for each pixel \(p_{ij}\). In practice, the effect of a
single pixel on the classification is negligible, so the masking of
larger regions is required to produce an interpretable map. 

Here, we employ two strategies to test occlusion sensitivity.
Firstly, we create annular masks with a width of 5 pixels and a radius varying
between 0 and 50 pixels. An annulus is chosen since the sources,
particularly strong lens systems, vary systematically with distance from
the centre of the lens (lens light decreases, and lensed images appear
around the Einstein radius of the system). Within the masks, the input
pixels are set to zero. We therefore test the sensitivity to information
in the image as a function of radius from the centre of the (real or
potential) lensing galaxy.

We also test the sensitivity to occlusion by a disk of a radius between 0 and 35 
pixels. As the disk grows in size and more information is removed from 
the images, we are able to test the sensitivity to the complete removal of information
within a certain radius $r$. From this we can quantify the amount of lensing
information present outside $r$; that is, we can test how sensitive the network
is to information inside $r$ being present. We can compare this directly to the 
Grad-CAM saliency maps from Figure~\ref{fig:saliencymaps}.

\subsubsection{Results}

\begin{figure*}
\hypertarget{fig:occlude}{%
\centering
\includegraphics[width=0.95\textwidth,keepaspectratio]{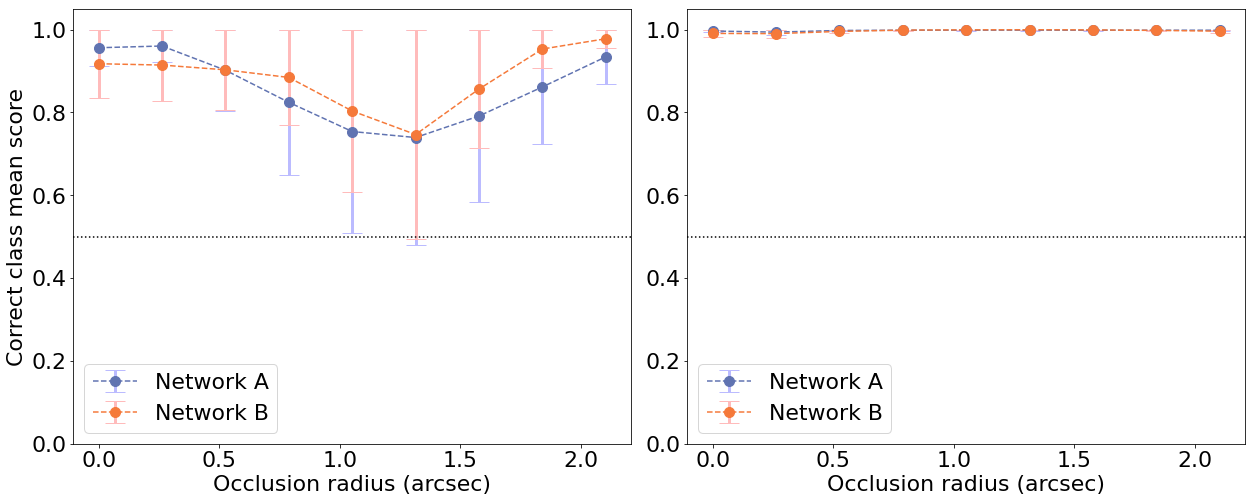}
\caption{Effect on the network scores of the occlusion of an annulus of
pixels with the given radius in arcsec (1 pixel = .263 arcsec).
\textbf{Left:} Effect on the score of simulated lenses, showing the
greatest effect at \(\sim 1.3-2.1 arcsec\), consistent with a typical
detectable strong lens. \textbf{Right}: Effect on the scores of
non-lenses. The networks are not sensitive to lack of radial information
at any radius for a non-lens.}\label{fig:occlude}
}

\end{figure*}

The experiment with ring occlusion provides a quantitative measure of some of the
morphological features learned by the network. In
Figure~\ref{fig:occlude} (right panel), we see that for both networks
the effect of occlusion (at any radius) does not affect the ability to
pick a non-lens galaxy. However, it has an impact on the scores of
lenses: at a radius of 1.3 arcsec (so, information between 1.3 and 2.1
arcsec is missing) this fact is most pronounced, with accuracy dropping
to \(\sim 75\%\). This corresponds with the Einstein radii of the
typical, and most common, lenses simulated and discovered in the survey.
Accuracy is decreased by more than 10\% across the range 0.5-1.7 arcsec.
This confirms the networks have learned that the presence of lensed
source flux at these radii is a key indicator of a strong lens, as
expected. The lack of any uncertainty introduced into the non-lens
scores when information is subtracted confirms this intuition. The fact
that the effect is maximum in the expected range indicates that there is
no concerning bias in the training set.

\begin{figure*}
\hypertarget{fig:occludedisk}{%
\centering
\includegraphics[width=0.95\textwidth,keepaspectratio]{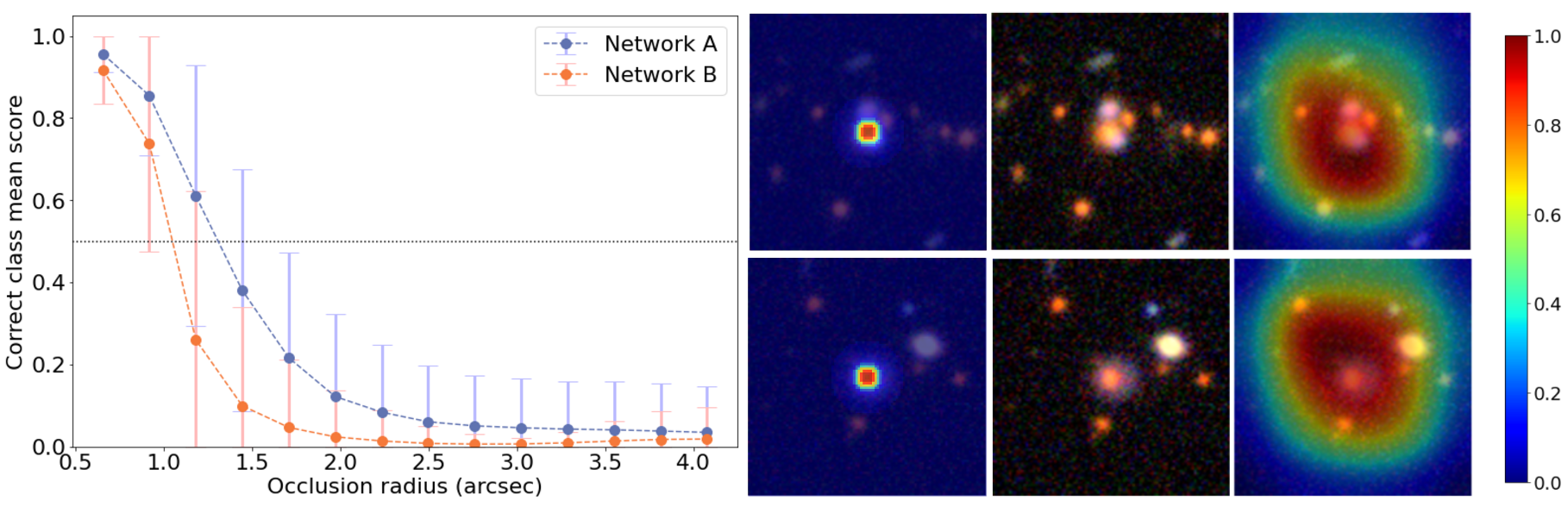}
\caption{Effect on the network scores for simulated strong lenses of 
the occlusion of a disk of pixels with the given radius in arcsec.
\textbf{Left:} 
For Network A, by radius 1.5 arcsec the majority of lenses are classified
as non lenses. For Network B, the decline is faster, at just over 1 arcsec.
\textbf{Right:}
    A comparison between the saliency maps produced using the Grad-CAM
algorithm (right panel) and the importance of radial information from the
    disk occlusion sensitivity analysis (left panel). The color scale represents both the
Grad-CAM salience and the information content by radius for Network A. Grad-CAM indicates
information within several arcsec is highly salient for these two lenses; 
our analysis indicates that 90\% of the relevant information lies within
2 arcsec for our test set.
}\label{fig:occludedisk}
}
\end{figure*}

In Figure~\ref{fig:occludedisk}, left, we see how the certainty that Networks A and B
have in identifying a lens drops as a function of disk occlusion radius. 
For network A, the mean score for lenses has dropped to .85 by a radius of .9 arcsec
and .12 by 2 arcsec. If we take this information as an approximation of 
the lensing information content as a function of $r$,  we see that $\sim85\%$ of the
information used by Network A lies within a radius of 1 arcsec (over the
supplied test set). We can compare this to the Grad-CAM test from 
Figure~\ref{fig:saliencymaps}. On the right of the figure we contrast
the Grad-CAM results for two lenses with this sensitivity map result. The 
Grad-CAM map corresponds to the individual input, while the sensitivity map result
is statistical; we can nevertheless see that the information content is more tightly
constrained to the central regions than the Grad-CAM maps may indicate.

For both networks, the accuracy on non-lenses is not affected by disk occlusion;
removing central flux never results in a decreased certainty that an object
is a non-lens.

\hypertarget{noise-1}{%
\subsection{Noise}\label{noise-1}}

\subsubsection{Test performed}

The addition of Gaussian noise enables us to probe the sensitivity of
the model to signal-to-noise; what is the threshold beyond which the
network can no longer reliably distinguish a lens, and how does this
compare to a human expert? We interrogate this through the addition of
increasing amounts of Gaussian noise to the images. The test set images
are normalized such that the flux in each band to a mean of zero and a
standard deviation of one, i.e.~\(X' = (X-\mu)/\sigma\), and then apply
Gaussian noise, parameterized by a standard deviation between 0 and 30
to each pixel in an image. This corresponds to to a typical change in
mean signal-to-noise ratio per pixel from 20 down to to \(\sim 0.5\)
across the test set. We test the sensitivity to the magnitude of this
Gaussian noise.

\subsubsection{Results}

\begin{figure*}
\hypertarget{fig:noise}{%
\centering
\includegraphics[width=0.95\textwidth,keepaspectratio]{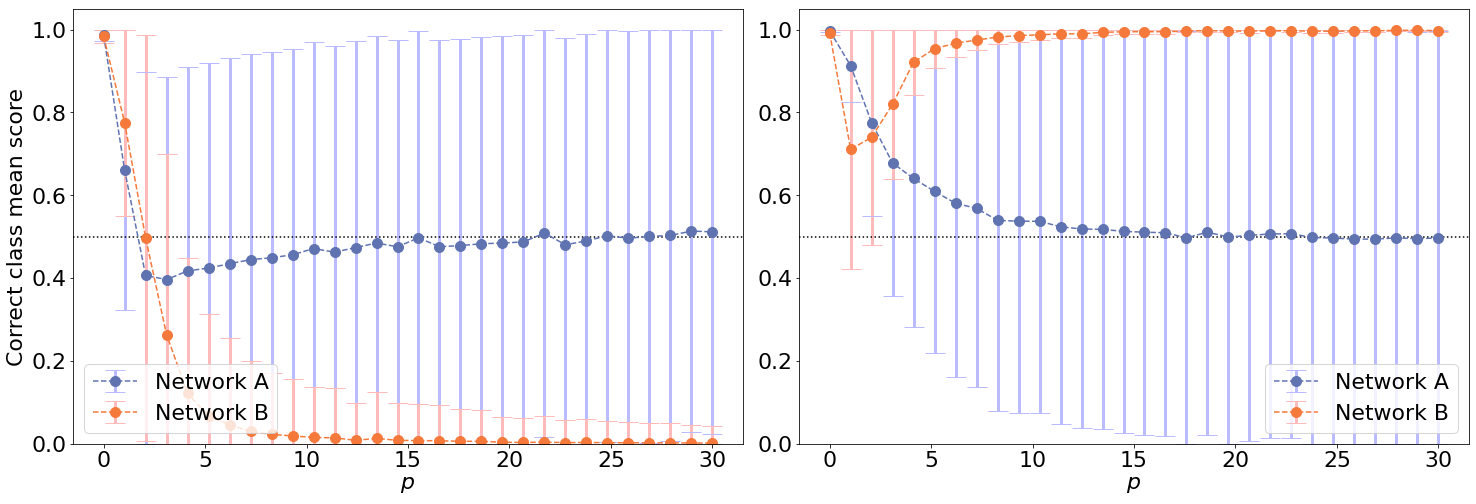}
\caption{The effects of added noise. Here, the parameter \(p\)
corresponds to the magnitude of Gaussian noise, corresponding to a
typical decrease in signal-to-noise to \(\sim 0.5\) when \(p=30\).
\textbf{Left}: The effect of noise on the scores of simulated lenses.
Two different strategies are evident. While Network A becomes more
uncertain as noise is added, Network B becomes more certain the example
is not a lens. \textbf{Right}: The same pattern is evident for the
non-lenses. Network A becomes more uncertain (mean of 0.5, high
scatter), whereas Network B becomes certain the examples are not
lenses.}\label{fig:noise}
}

\end{figure*}

The addition of noise to the test set decreased the accuracy of the
network as expected. However, the two networks behaved differently,
shedding some light on the different features learned from the two
training sets. For network A, the addition of noise leads to confusion -
both simulated lenses and non lens galaxies are scored as uncertain (0.5) as
the images get noiser (Figure~\ref{fig:noise}, left). However, network B
scores simulated lenses as more likely to be non-lenses as the noise increases and
the real non-lens galaxies becomes \emph{more} certain as noise
increases. For network A, if it cannot discern lens morphology, it
becomes uncertain; for network B, it becomes certain that the object is
not a strong lens. For a balanced test set, the overall accuracy will be
the same in both cases, but in practice the network B strategy is likely
to lead to fewer false positives, albeit at the cost of completeness, in
lower signal-to-noise regimes. The fact that strong lenses are very rare
also points to the Network B strategy being more robust; in the absence
of clear lensing features, the candidate would ideally be rejected,
rather than given an uncertain score.

\hypertarget{effects-on-real-images-of-gravitational-lenses}{%
\subsection{Effects on real images of gravitational
lenses}\label{effects-on-real-images-of-gravitational-lenses}}

\begin{figure*}
\hypertarget{fig:real_lenses}{%
\centering
\includegraphics[width=0.95\textwidth,keepaspectratio]{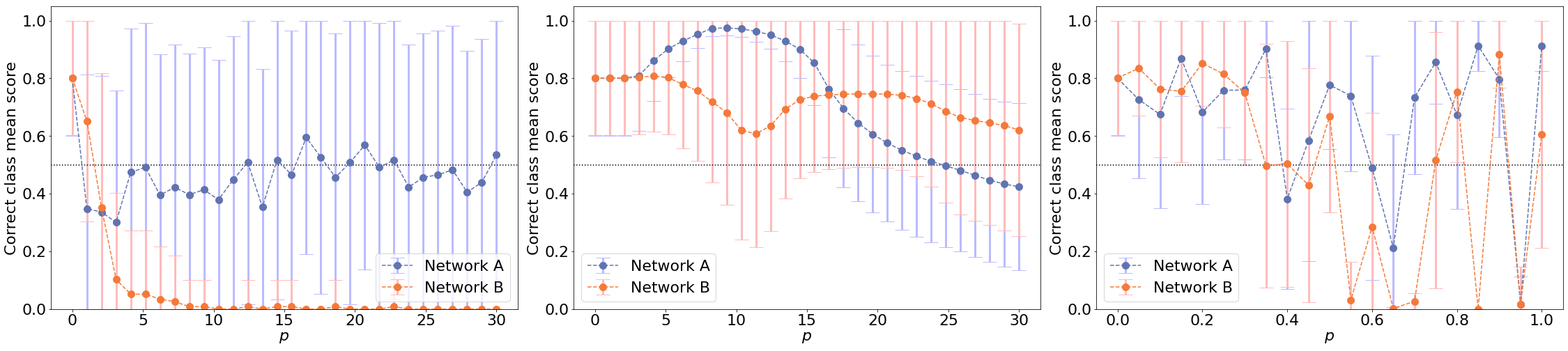}
\caption{Effects of several sensitivity tests on the two networks using
a test set consisting of 116 images of strong lens candidates from DES.
\textbf{Left}: The effect of adding noise. \textbf{Middle}: The effect
of convolution with a Gaussian blur. \textbf{Right}: The effect of
color jitter. For the blur and noise addition, the results are
consistent with what we see on the simulations, with Network A's
apparent increase in performance for small blurs explained by a
corresponding jump in false positives under this transformation. The
effect of color jitter is strong, but due to the small size of the test
set and the large amount of scatter, it is difficut to quantify beyond
noting the drastic impact on accuracy.}\label{fig:real_lenses}
}

\end{figure*}

Several of our perturber-based sensitivity tests were applied to a small
test set containing images of real strong lenses.
Figure~\ref{fig:real_lenses} shows the results of applying the noise,
blur and color jitter tests as described previously. The noise and blur
tests are similar to the tests performed on simulations, with the
interesting feature of an increase in blurred accuracy for Network A up
to a point, then the expected decrease. This does not imply that
blurring the inputs would improve performance in reality, since we see
that for non-lenses the blurring decreases the correct score---the false
positive rate would increase dramatically in this case. Color jitter
can also be seen to have a substantial impact on the networks' accuracy;
the scatter evident in the figure is a function of the small size of the
test set.

\subsection{Limitations of the method}

The sensitivity probe is a relatively simple method and has several
limitations. As presented here in the lens-finding case, 
the properties to be tested must be known \textit{a priori};
the method does not, in an unsupervised way, segregate the test set
in a way that allows discovery of hidden biases, for instance
by automatically segregating the test set into groups of over- and
under-performing objects. However, in astronomy we often do have 
a catalog of properties associated with the objects of interest: 
photometric magnitudes, flux errors, photometric and 
spectroscopic redshifts, colors, etc. In these cases the method
is both useful and readily applied.  
Errors in these catalog properties 
will affect the sensitivity probe
analysis, however due to its statistical nature it can still
provide interpretable results even with noisily-labelled
data. Even dividing the test set into as few as two or three 
bins would enable strong biases to be visualised.

Correlations between biases are also not analyzed. Since the model 
produces a single score for each object, the correlations in score
quality for the lensing application simply reflect the correlations
between galaxy and observational properties. However, segregating
the test set into bins in two or more dimensions (i.e. by two or more
properties) would be possible and for some applications such a `performance
surface' may offer useful insights.

\hypertarget{results-summary}{%
\subsection{Results summary}\label{results-summary}}

We tested the sensitivity of two trained neural networks to the
\emph{g}-band source magnitude, \emph{r}-band lens magnitude, Einstein
radius and PSF of simulated strong lenses, and also to the addition of
noise, a Gaussian blur, and random rescaling of different bands in the
images (color jitter). The networks were highly sensitive to the color
jitter and PSF, showed some sensitivity to the galaxy magnitudes, and
were insensitive to the simulated Einstein radius. The addition of noise
and blur, which removed information from the images also caused a smooth
decrease in network performance. From this we conclude in particular
that the PSF is likely to be a significant weakness in the training set
design, that the networks may benefit from a wider range of colors in
the training sets, and that the networks are probably unreliable where
the source is too faint (\(g > 21.5\)) or the lens too bright
(\(r < 19\)). These results are summarized in
Table~\ref{tbl:results_2col}.

The networks tested in this work were successful in facilitating the
discovery of \({\sim}500\) high-quality strong lens candidates in DES
imaging. The completeness of that search is difficult to ascertain; it
can be estimated using simulations \citep[as
per][]{collettModelIndependentDeterminationH02019} or perhaps more
certainly when other techniques (such as mining spectroscopic data) are
employed exhaustively to a significant subset of the survey area in
future. We cannot easily quantify the properties of the sources missed
or falsely rejected, nor can we easily understand in many cases why
false positives are scored highly. The use of a sensitivity probe has
allowed us to probe regions of parameter space where the performance of
the networks degrades, putting some constraints on a selection function
in terms of signal-to-noise, galaxy magnitudes, and PSF/seeing. This will
enable the refinement of some estimates of the number of discoverable
lenses in future surveys, and also provide a benchmark to which the next
generation of lens-finders can be compared.

\hypertarget{tbl:results_2col}{}
\begin{table*}
\centering

\caption{\label{tbl:results_2col}A summary of the tests performed and
conclusions drawn from the sensitivity probe.}

\begin{tabular}{@{}lllll@{}}

\toprule

\textbf{Test} & \textbf{Parameter tested} & \textbf{Range} & \textbf{Significance} & \textbf{Comment} \\ \midrule
Color jitter & Band fluxes scaled to alter color & 0-110\% & high & Greater variation of colors may assist \\
& & & &   training robustness.  \\
Einstein radius & Radius in arcsec & 1.9-2.75 & none & Network functions well across a  \\
& & & & large range.  \\
PSF & Scaling factor from fiducial PSF & 0.6-3 & high & Network overly sensitive to simulated PSF, \\
 & & & & greater range indicated for training. \\
Occlusion & Radius of occluded annulus & 0-2 arcsec & high & Network learns to rely on information \\
 & & & & at typical Einstein radius (1.3 arcsec).  \\
Noise & Gaussian noise added & $20 \lessapprox \textrm{S/N}  \lessapprox 0.5$ & med & Performance degrades smoothly; \\
 & & & & different strategy between two networks. \\
Blur & Convolution with Gaussian kernel & $0 < \textrm{kernel} < 7.9'' $ & med & Performance degrades $> 1.3$ arcsec. \\
Lens mag & $r$-band magnitude of lens galaxy & 21-23 & low-med & Network A not sensitive across this range. \\
 & & & & Network B mean score dropped to 0.8 \\
 & & & & by $r=20.5$. \\
Source mag & $g$-band magnitude of lens galaxy & 22-24 & med-high & Network A robust until $g > 21.5$; Network \\
& & & &   R until $g \sim 21$. \\
 \bottomrule

\end{tabular}

\end{table*}

\hypertarget{sec:conclusion}{%
\section{Conclusion}\label{sec:conclusion}}

The use of deep neural networks in astronomy is growing exponentially,
playing an increasing role in many subject areas. Understanding the
biases and weaknesses of deep learning models is important to properly
place the results in context and use them in downstream scientific
analysis. In this article we use a sensitivity probe---which tests how
the accuracy of a neural network varies as a function of a specified
property of the input data---to probe two neural networks trained to
recognise images of gravitational lenses in the Dark Energy Survey. We
test the sensitivity of the networks to seeing (a Gaussian blur); a
simulated point spread function; color; noise; occlusion of an annulus
of pixels; the lens \(r\)-band magnitude; the lensed source's \(g\)-band
magnitude; and the Einstein radius of (simulated) strong lenses. We find
that the networks are highly sensitive to color, indicating that they
have learned that the colors of a typical strong lens (a red elliptical
lens with a blue star-forming lensed source) are significant; this may
indicate that an atypical lens, such as one with a red source, would be
scored low. The network is also sensitive to the simulated PSF, with
performance degrading smoothly as the PSF broadens from the fiducial
value used to train the network, but also degrades rapidly as the PSF
improves, indicating that the network is overly sensitive to the
simulated PSF and should ideally be trained with a broader range of
simulations to address this weakness. Response to noise, blur, and the
lens and source magnitudes degrades smoothly as expected, however at a
g-band magnitude of \(\sim 21.5\) the performance is observed to degrade
significantly, allowing us to constrain the selection function of the
lens-finder. The network is not sensitive to the Einstein radius of an
input lens image, reporting similar accuracy across a range from 1.0 to
2.7 arcsec. We tested the networks' sensitivity to the occlusion
(zeroing-out) of a ring three pixels wide; the network was most
sensitive to this effect when the radius of the occlusion ring was at 5
pixels, equivalent to 1.3 arcsec. This is typical of a strong lens
detected in the DES imaging and confirms that the network has learned
this is a highly significant region of the image. These insights
highlight potential utility of a sensitivity probe in finding
weaknesses in deep learning-based algorithms and their training sets.

Future work could assist the next generation of lens-finders by
quantifying the significance of assumptions underlying the simulations
which for now compose the only feasible training set. The Sersic index
of the simulated source, the presence of multiple lensed sources,
clumpiness in the sources, and more complicated mass distributions in
the lens plane could all be tested.

Although the sensitivity probe method relies only on the collection of 
performance statistics calculated from the outputs of the networks,
and is therefore simple to implement, it has not been systematically
applied before in lens-finding or other DNN-driven astronomical applications.
A sensitivity probe such as that presented here could prove useful more
generally in future deep-learning based experiments by informing
experimental design and identifying weaknesses in training sets that
could be remedied before scientific application. The sensitivity probe
also has the potential to help us understand the features a network has
learned are the most salient in a particular context, from which we may
potentially be able to ascribe astrophysical significance. Combined with
new techniques for estimating the uncertainties in ANN outputs such as
Bayesian neural networks, the chief weakness of deep learning approaches
to science---the lack of interpretability---may be significantly
ameliorated.

\hypertarget{acknowledgements}{%
\section{Acknowledgements}\label{acknowledgements}}

The author acknowledges support from Karl Glazebrook's Australian
Research Council Laureate Fellowship FL180100060. This research was
supported by use of the Nectar Research Cloud and by Swinburne
University of Technology. The Nectar Research Cloud is a collaborative
Australian research platform supported by the NCRIS-funded Australian
Research Data Commons (ARDC).

\bibliographystyle{mnras}
\renewcommand\refname{References}
\bibliography{probe}

\begin{thebibliography}{}
\makeatletter
\relax
\def\mn@urlcharsother{\let\do\@makeother \do\$\do\&\do\#\do\^\do\_\do\%\do\~}
\def\mn@doi{\begingroup\mn@urlcharsother \@ifnextchar [ {\mn@doi@}
  {\mn@doi@[]}}
\def\mn@doi@[#1]#2{\def\@tempa{#1}\ifx\@tempa\@empty \href
  {http://dx.doi.org/#2} {doi:#2}\else \href {http://dx.doi.org/#2} {#1}\fi
  \endgroup}
\def\mn@eprint#1#2{\mn@eprint@#1:#2::\@nil}
\def\mn@eprint@arXiv#1{\href {http://arxiv.org/abs/#1} {{\tt arXiv:#1}}}
\def\mn@eprint@dblp#1{\href {http://dblp.uni-trier.de/rec/bibtex/#1.xml}
  {dblp:#1}}
\def\mn@eprint@#1:#2:#3:#4\@nil{\def\@tempa {#1}\def\@tempb {#2}\def\@tempc
  {#3}\ifx \@tempc \@empty \let \@tempc \@tempb \let \@tempb \@tempa \fi \ifx
  \@tempb \@empty \def\@tempb {arXiv}\fi \@ifundefined
  {mn@eprint@\@tempb}{\@tempb:\@tempc}{\expandafter \expandafter \csname
  mn@eprint@\@tempb\endcsname \expandafter{\@tempc}}}

\bibitem[\protect\citeauthoryear{Binder, Montavon, Lapuschkin, M{\"u}ller  \&
  Samek}{Binder et~al.}{2016}]{binderLayerwiseRelevancePropagation2016}
Binder A.,  Montavon G.,  Lapuschkin S.,  M{\"u}ller K.-R.,   Samek W.,  2016,
  in International {{Conference}} on {{Artificial Neural Networks}}.
  {Springer}, pp 63--71

\bibitem[\protect\citeauthoryear{Birrer, Amara  \& Refregier}{Birrer
  et~al.}{2017}]{BirrerLensingsubstructurequantification2017}
Birrer S.,  Amara A.,   Refregier A.,  2017, \mn@doi [J. Cosmol. Astropart.
  Phys.] {10.1088/1475-7516/2017/05/037}, 2017, 037

\bibitem[\protect\citeauthoryear{Birrer et~al.,}{Birrer
  et~al.}{2019}]{birrerH0LiCOWIXCosmographic2019}
Birrer S.,  et~al., 2019, \mn@doi [Monthly Notices of the Royal Astronomical
  Society] {10.1093/mnras/stz200}, 484, 4726

\bibitem[\protect\citeauthoryear{Bonvin et~al.,}{Bonvin
  et~al.}{2016}]{BonvinH0LiCOWNewCOSMOGRAIL2016}
Bonvin V.,  et~al., 2016, \mn@doi [MNRAS] {10.1093/mnras/stw3006}, p. stw3006

\bibitem[\protect\citeauthoryear{Brault \& Gavazzi}{Brault \&
  Gavazzi}{2015}]{braultExtensiveLightProfile2015}
Brault F.,  Gavazzi R.,  2015, \mn@doi [Astronomy \& Astrophysics]
  {10.1051/0004-6361/201425275}, 577, A85

\bibitem[\protect\citeauthoryear{Chan, Suyu, Chiueh, More, Marshall, Coupon,
  Oguri  \& Price}{Chan
  et~al.}{2015}]{ChanChitahStronggravitationallensHunter2015}
Chan J. H.~H.,  Suyu S.~H.,  Chiueh T.,  More A.,  Marshall P.~J.,  Coupon J.,
  Oguri M.,   Price P.,  2015, \mn@doi [The Astrophysical Journal]
  {10.1088/0004-637X/807/2/138}, 807

\bibitem[\protect\citeauthoryear{Collett}{Collett}{2015}]{collett_population_2015}
Collett T.~E.,  2015, \mn@doi [ApJ] {10.1088/0004-637X/811/1/20}, 811, 20

\bibitem[\protect\citeauthoryear{Collett \& Smith}{Collett \&
  Smith}{2020}]{collettTripleRolloverThird2020}
Collett T.~E.,  Smith R.~J.,  2020, arXiv:2004.00649 [astro-ph]

\bibitem[\protect\citeauthoryear{Collett, Montanari  \&
  R{\"a}s{\"a}nen}{Collett
  et~al.}{2019}]{collettModelIndependentDeterminationH02019}
Collett T.,  Montanari F.,   R{\"a}s{\"a}nen S.,  2019, \mn@doi [Physical
  Review Letters] {10.1103/PhysRevLett.123.231101}, 123, 231101

\bibitem[\protect\citeauthoryear{{Dark Energy Survey Collaboration}
  et~al.,}{{Dark Energy Survey Collaboration}
  et~al.}{2016}]{darkenergysurveycollaborationDarkEnergySurvey2016}
{Dark Energy Survey Collaboration} et~al., 2016, \mn@doi [MNRAS]
  {10.1093/mnras/stw641}, 460, 1270

\bibitem[\protect\citeauthoryear{Devlin, Chang, Lee  \& Toutanova}{Devlin
  et~al.}{2019}]{devlinBERTPretrainingDeep2019}
Devlin J.,  Chang M.-W.,  Lee K.,   Toutanova K.,  2019, in Proceedings of the
  2019 {{Conference}} of the {{North American Chapter}} of the {{Association}}
  for {{Computational Linguistics}}: {{Human Language Technologies}},
  {{Volume}} 1 ({{Long}} and {{Short Papers}}). {Association for Computational
  Linguistics}, {Minneapolis, Minnesota}, pp 4171--4186,
  \mn@doi{10.18653/v1/N19-1423}, \url
  {https://www.aclweb.org/anthology/N19-1423}

\bibitem[\protect\citeauthoryear{Dieleman, Willett  \& Dambre}{Dieleman
  et~al.}{2015}]{dieleman_rotation-invariant_2015}
Dieleman S.,  Willett K.~W.,   Dambre J.,  2015, \mn@doi [Monthly Notices of
  the Royal Astronomical Society] {10.1093/mnras/stv632}, 450, 1441

\bibitem[\protect\citeauthoryear{Eriksen et~al.,}{Eriksen
  et~al.}{2020}]{eriksenPAUSurveyPhotometric2020}
Eriksen M.,  et~al., 2020, arXiv:2004.07979 [astro-ph]

\bibitem[\protect\citeauthoryear{Fluke \& Jacobs}{Fluke \&
  Jacobs}{2020}]{flukeSurveyingReachMaturity2020}
Fluke C.~J.,  Jacobs C.,  2020, \mn@doi [WIREs Data Mining and Knowledge
  Discovery] {10.1002/widm.1349}, 10, e1349

\bibitem[\protect\citeauthoryear{Fukushima}{Fukushima}{1980}]{fukushimaNeocognitronSelforganizingNeural1980}
Fukushima K.,  1980, \mn@doi [Biol. Cybernetics] {10.1007/BF00344251}, 36, 193

\bibitem[\protect\citeauthoryear{Gavazzi, Marshall, Treu  \&
  Sonnenfeld}{Gavazzi et~al.}{2014}]{gavazzi_ringfinder:_2014}
Gavazzi R.,  Marshall P.~J.,  Treu T.,   Sonnenfeld A.,  2014, \mn@doi [ApJ]
  {10.1088/0004-637X/785/2/144}, 785, 144

\bibitem[\protect\citeauthoryear{George \& Huerta}{George \&
  Huerta}{2018}]{georgeDeepLearningRealtime2018}
George D.,  Huerta E.~A.,  2018, \mn@doi [Physics Letters B]
  {10.1016/j.physletb.2017.12.053}, 778, 64

\bibitem[\protect\citeauthoryear{Gilpin, Bau, Yuan, Bajwa, Specter  \&
  Kagal}{Gilpin et~al.}{2018}]{gilpinExplainingExplanationsOverview2018}
Gilpin L.~H.,  Bau D.,  Yuan B.~Z.,  Bajwa A.,  Specter M.,   Kagal L.,  2018,
  in 2018 {{IEEE}} 5th {{International Conference}} on {{Data Science}} and
  {{Advanced Analytics}} ({{DSAA}}). pp 80--89,
  \mn@doi{10.1109/DSAA.2018.00018}

\bibitem[\protect\citeauthoryear{Hoyle}{Hoyle}{2016}]{hoyleMeasuringPhotometricRedshifts2016}
Hoyle B.,  2016, \mn@doi [Astronomy and Computing]
  {10.1016/j.ascom.2016.03.006}, 16, 34

\bibitem[\protect\citeauthoryear{Ilbert et~al.,}{Ilbert
  et~al.}{2009}]{ilbertCosmosPhotometricRedshifts2009}
Ilbert O.,  et~al., 2009, \mn@doi [ApJ] {10.1088/0004-637X/690/2/1236}, 690,
  1236

\bibitem[\protect\citeauthoryear{Iten, Metger, Wilming, {del Rio}  \&
  Renner}{Iten et~al.}{2020}]{itenDiscoveringPhysicalConcepts2020}
Iten R.,  Metger T.,  Wilming H.,  {del Rio} L.,   Renner R.,  2020, \mn@doi
  [Phys. Rev. Lett.] {10.1103/PhysRevLett.124.010508}, 124, 010508

\bibitem[\protect\citeauthoryear{Ivezi{\'c} et~al.,}{Ivezi{\'c}
  et~al.}{2019}]{ivezicLSSTScienceDrivers2019}
Ivezi{\'c} {\v Z}.,  et~al., 2019, \mn@doi [ApJ] {10.3847/1538-4357/ab042c},
  873, 111

\bibitem[\protect\citeauthoryear{Jacobs}{Jacobs}{2020}]{jacobsSensieProbingSensitivity2020}
Jacobs C.,  2020, \mn@doi [Journal of Open Source Software]
  {10.21105/joss.02180}, 5, 2180

\bibitem[\protect\citeauthoryear{Jacobs, Glazebrook, Collett, More  \&
  McCarthy}{Jacobs et~al.}{2017}]{jacobsFindingStrongLenses2017}
Jacobs C.,  Glazebrook K.,  Collett T.,  More A.,   McCarthy C.,  2017, \mn@doi
  [Mon Not R Astron Soc] {10.1093/mnras/stx1492}, 471, 167

\bibitem[\protect\citeauthoryear{Jacobs et~al.,}{Jacobs
  et~al.}{2019a}]{jacobsExtendedCatalogGalaxy2019}
Jacobs C.,  et~al., 2019a, \mn@doi [ApJS] {10.3847/1538-4365/ab26b6}, 243, 17

\bibitem[\protect\citeauthoryear{Jacobs et~al.,}{Jacobs
  et~al.}{2019b}]{jacobsFindingHighredshiftStrong2019}
Jacobs C.,  et~al., 2019b, \mn@doi [Mon Not R Astron Soc]
  {10.1093/mnras/stz272}, 484, 5330

\bibitem[\protect\citeauthoryear{Jones, Stark  \& Ellis}{Jones
  et~al.}{2018}]{jonesDustWindComposition2018}
Jones T.,  Stark D.~P.,   Ellis R.~S.,  2018, \mn@doi [The Astrophysical
  Journal] {10.3847/1538-4357/aad37f}, 863, 191

\bibitem[\protect\citeauthoryear{Kim \& Brunner}{Kim \&
  Brunner}{2016}]{kimStargalaxyClassificationUsing2016}
Kim E.~J.,  Brunner R.~J.,  2016, arXiv:1608.04369 [astro-ph]

\bibitem[\protect\citeauthoryear{Kindermans, Sch{\"u}tt, Alber, M{\"u}ller,
  Erhan, Kim  \& D{\"a}hne}{Kindermans
  et~al.}{2017}]{kindermansLearningHowExplain2017}
Kindermans P.-J.,  Sch{\"u}tt K.~T.,  Alber M.,  M{\"u}ller K.-R.,  Erhan D.,
  Kim B.,   D{\"a}hne S.,  2017, arXiv:1705.05598 [cs, stat]

\bibitem[\protect\citeauthoryear{Kremer, {Stensbo-Smidt}, Gieseke,
  Steenstrup~Pedersen  \& Igel}{Kremer et~al.}{2017}]{kremerBigUniverseBig2017}
Kremer J.,  {Stensbo-Smidt} K.,  Gieseke F.,  Steenstrup~Pedersen K.,   Igel
  C.,  2017, preprint, 1704, arXiv:1704.04650

\bibitem[\protect\citeauthoryear{Krizhevsky, Sutskever  \& Hinton}{Krizhevsky
  et~al.}{2012}]{krizhevskyImageNetClassificationDeep2012}
Krizhevsky A.,  Sutskever I.,   Hinton G.~E.,  2012, in Pereira F.,  Burges C.
  J.~C.,  Bottou L.,   Weinberger K.~Q.,  eds, , Advances in {{Neural
  Information Processing Systems}} 25.
{Curran Associates, Inc.}, pp 1097--1105, \url
  {http://papers.nips.cc/paper/4824-imagenet-classification-with-deep-convolutional-neural-networks.pdf}

\bibitem[\protect\citeauthoryear{LeCun, Boser, Denker, Henderson, Howard,
  Hubbard  \& Jackel}{LeCun
  et~al.}{1989}]{LeCunBackpropagationAppliedHandwritten1989}
LeCun Y.,  Boser B.,  Denker J.~S.,  Henderson D.,  Howard R.~E.,  Hubbard W.,
   Jackel L.~D.,  1989, \mn@doi [Neural Computation]
  {10.1162/neco.1989.1.4.541}, 1, 541

\bibitem[\protect\citeauthoryear{LeCun, Kavukcuoglu, Farabet  et~al.}{LeCun
  et~al.}{2010}]{lecunConvolutionalNetworksApplications2010}
LeCun Y.,  Kavukcuoglu K.,  Farabet C.,   et~al., 2010, in {{ISCAS}}. pp
  253--256, \url
  {http://research2.fit.edu/ice/sites/default/files/Convolutional%20networks%20and%20applications%20in%20vision_0.pdf}

\bibitem[\protect\citeauthoryear{LeCun, Bengio  \& Hinton}{LeCun
  et~al.}{2015}]{lecunDeepLearning2015}
LeCun Y.,  Bengio Y.,   Hinton G.,  2015, \mn@doi [Nature]
  {10.1038/nature14539}, 521, 436

\bibitem[\protect\citeauthoryear{Li, Frenk, Cole, Gao, Bose  \& Hellwing}{Li
  et~al.}{2016}]{LiConstraintsidentitydark2016}
Li R.,  Frenk C.~S.,  Cole S.,  Gao L.,  Bose S.,   Hellwing W.~A.,  2016,
  \mn@doi [Monthly Notices of the Royal Astronomical Society]
  {10.1093/mnras/stw939}, 460

\bibitem[\protect\citeauthoryear{Marshall, Hogg, Moustakas, Fassnacht, Brada{\v
  c}, {Tim Schrabback}  \& Blandford}{Marshall
  et~al.}{2009}]{marshall_automated_2009}
Marshall P.~J.,  Hogg D.~W.,  Moustakas L.~A.,  Fassnacht C.~D.,  Brada{\v c}
  M.,  {Tim Schrabback}  Blandford R.~D.,  2009, \mn@doi [ApJ]
  {10.1088/0004-637X/694/2/924}, 694, 924

\bibitem[\protect\citeauthoryear{Marshall et~al.,}{Marshall
  et~al.}{2016}]{marshallSPACEWARPSCrowdsourcing2016}
Marshall P.~J.,  et~al., 2016, \mn@doi [Monthly Notices of the Royal
  Astronomical Society] {10.1093/mnras/stv2009}, 455

\bibitem[\protect\citeauthoryear{Ntampaka, Eisenstein, Yuan  \&
  Garrison}{Ntampaka et~al.}{2020}]{ntampakaHybridDeepLearning2020}
Ntampaka M.,  Eisenstein D.~J.,  Yuan S.,   Garrison L.~H.,  2020, \mn@doi
  [ApJ] {10.3847/1538-4357/ab5f5e}, 889, 151

\bibitem[\protect\citeauthoryear{Oguri, Rusu  \& Falco}{Oguri
  et~al.}{2014}]{Oguristellardarkmatter2014}
Oguri M.,  Rusu C.~E.,   Falco E.~E.,  2014, \mn@doi [Monthly Notices of the
  Royal Astronomical Society] {10.1093/mnras/stu106}, 439

\bibitem[\protect\citeauthoryear{Petrillo et~al.,}{Petrillo
  et~al.}{2019}]{petrilloLinKSDiscoveringGalaxyscale2019}
Petrillo C.~E.,  et~al., 2019, MNRAS, 484, 3879

\bibitem[\protect\citeauthoryear{Rosenblatt}{Rosenblatt}{1957}]{rosenblattPerceptronaPerceivingRecognizing1957}
Rosenblatt F.,  1957, Cornell Aeronautical Lab

\bibitem[\protect\citeauthoryear{Russakovsky et~al.,}{Russakovsky
  et~al.}{2015}]{russakovskyImageNetLargeScale2015}
Russakovsky O.,  et~al., 2015, \mn@doi [Int J Comput Vis]
  {10.1007/s11263-015-0816-y}, 115, 211

\bibitem[\protect\citeauthoryear{Schmidhuber}{Schmidhuber}{2015}]{schmidhuberDeepLearningNeural2015}
Schmidhuber J.,  2015, \mn@doi [Neural Networks]
  {10.1016/j.neunet.2014.09.003}, 61, 85

\bibitem[\protect\citeauthoryear{Seidel \& Bartelmann}{Seidel \&
  Bartelmann}{2007}]{seidel_arcfinder:_2007}
Seidel G.,  Bartelmann M.,  2007, \mn@doi [Astronomy \& Astrophysics]
  {10.1051/0004-6361:20066097}, 472, 12

\bibitem[\protect\citeauthoryear{Selvaraju, Cogswell, Das, Vedantam, Parikh  \&
  Batra}{Selvaraju et~al.}{2017}]{selvarajuGradCAMVisualExplanations2017}
Selvaraju R.~R.,  Cogswell M.,  Das A.,  Vedantam R.,  Parikh D.,   Batra D.,
  2017, in Proceedings of the {{IEEE International Conference}} on {{Computer
  Vision}}. pp 618--626, \url
  {http://openaccess.thecvf.com/content_iccv_2017/html/Selvaraju_Grad-CAM_Visual_Explanations_ICCV_2017_paper.html}

\bibitem[\protect\citeauthoryear{{Sevilla-Noarbe} et~al.,}{{Sevilla-Noarbe}
  et~al.}{2021}]{sevilla-noarbeDarkEnergySurvey2021}
{Sevilla-Noarbe} I.,  et~al., 2021, \mn@doi [ApJS] {10.3847/1538-4365/abeb66},
  254, 24

\bibitem[\protect\citeauthoryear{Sharma, Kembhavi, Kembhavi, Sivarani, Abraham
  \& Vaghmare}{Sharma et~al.}{2019}]{sharmaApplicationConvolutionalNeural2019}
Sharma K.,  Kembhavi A.,  Kembhavi A.,  Sivarani T.,  Abraham S.,   Vaghmare
  K.,  2019, arXiv e-prints, p. arXiv:1909.05459

\bibitem[\protect\citeauthoryear{Simonyan \& Zisserman}{Simonyan \&
  Zisserman}{2014}]{simonyanVeryDeepConvolutional2014}
Simonyan K.,  Zisserman A.,  2014, arXiv:1409.1556 [cs]

\bibitem[\protect\citeauthoryear{Simonyan, Vedaldi  \& Zisserman}{Simonyan
  et~al.}{2013}]{simonyanDeepConvolutionalNetworks2013}
Simonyan K.,  Vedaldi A.,   Zisserman A.,  2013, arXiv preprint arXiv:1312.6034

\bibitem[\protect\citeauthoryear{Smilkov, Thorat, Kim, Vi{\'e}gas  \&
  Wattenberg}{Smilkov et~al.}{2017}]{smilkovSmoothGradRemovingNoise2017}
Smilkov D.,  Thorat N.,  Kim B.,  Vi{\'e}gas F.,   Wattenberg M.,  2017,
  arXiv:1706.03825 [cs, stat]

\bibitem[\protect\citeauthoryear{Sonnenfeld, Treu, Gavazzi, Suyu, Marshall,
  Auger  \& Nipoti}{Sonnenfeld
  et~al.}{2013}]{sonnenfeldSL2SGalaxyscaleLens2013}
Sonnenfeld A.,  Treu T.,  Gavazzi R.,  Suyu S.~H.,  Marshall P.~J.,  Auger
  M.~W.,   Nipoti C.,  2013, \mn@doi [The Astrophysical Journal]
  {10.1088/0004-637X/777/2/98}, 777, 98

\bibitem[\protect\citeauthoryear{Sonnenfeld et~al.,}{Sonnenfeld
  et~al.}{2018}]{sonnenfeldSurveyGravitationallylensedObjects2018}
Sonnenfeld A.,  et~al., 2018, \mn@doi [Publ Astron Soc Jpn Nihon Tenmon Gakkai]
  {10.1093/pasj/psx062}, 70

\bibitem[\protect\citeauthoryear{Sonnenfeld et~al.,}{Sonnenfeld
  et~al.}{2020}]{sonnenfeldSurveyGravitationallylensedObjects2020}
Sonnenfeld A.,  et~al., 2020, arXiv e-prints, 2004, arXiv:2004.00634

\bibitem[\protect\citeauthoryear{Spilker}{Spilker}{2019}]{spilkerGalacticOutflowsHigh2019}
Spilker J.,  2019, \mn@doi [Proceedings of the International Astronomical
  Union] {10.1017/S1743921319009232}, 15, 187

\bibitem[\protect\citeauthoryear{Springenberg, Dosovitskiy, Brox  \&
  Riedmiller}{Springenberg
  et~al.}{2015}]{springenbergStrivingSimplicityAll2015}
Springenberg J.,  Dosovitskiy A.,  Brox T.,   Riedmiller M.,  2015, in {ICLR
  (workshop track)}. \url
  {https://lmb.informatik.uni-freiburg.de/Publications/2015/DB15a/}

\bibitem[\protect\citeauthoryear{Treu}{Treu}{2010}]{treu_strong_2010}
Treu T.,  2010, \mn@doi [Annual Review of Astronomy and Astrophysics]
  {10.1146/annurev-astro-081309-130924}, 48, 87

\bibitem[\protect\citeauthoryear{Voulodimos, Doulamis, Doulamis  \&
  Protopapadakis}{Voulodimos et~al.}{2018}]{voulodimosDeepLearningComputer2018}
Voulodimos A.,  Doulamis N.,  Doulamis A.,   Protopapadakis E.,  2018, Deep
  {{Learning}} for {{Computer Vision}}: {{A Brief Review}},
  \mn@doi{10.1155/2018/7068349}, \url
  {https://www.hindawi.com/journals/cin/2018/7068349/}

\bibitem[\protect\citeauthoryear{Walmsley et~al.,}{Walmsley
  et~al.}{2020}]{walmsleyGalaxyZooProbabilistic2020}
Walmsley M.,  et~al., 2020, \mn@doi [Mon Not R Astron Soc]
  {10.1093/mnras/stz2816}, 491, 1554

\bibitem[\protect\citeauthoryear{Wang, Xie, Zhang, Huang, Zhang  \& Liu}{Wang
  et~al.}{2020}]{wangLikelihoodfreeCosmologicalConstraints2020}
Wang Y.-C.,  Xie Y.-B.,  Zhang T.-J.,  Huang H.-C.,  Zhang T.,   Liu K.,  2020,
  arXiv e-prints, 2005, arXiv:2005.10628

\bibitem[\protect\citeauthoryear{Zeiler \& Fergus}{Zeiler \&
  Fergus}{2014}]{zeilerVisualizingUnderstandingConvolutional2014}
Zeiler M.~D.,  Fergus R.,  2014, in Fleet D.,  Pajdla T.,  Schiele B.,
  Tuytelaars T.,  eds, , Vol.~8689, Computer {{Vision}} \textendash{} {{ECCV}}
  2014.
{Springer International Publishing}, {Cham}, pp 818--833, \url
  {http://link.springer.com/10.1007/978-3-319-10590-1_53}

\bibitem[\protect\citeauthoryear{Zhang \& Zhao}{Zhang \&
  Zhao}{2015}]{zhangAstronomyBigData2015}
Zhang Y.,  Zhao Y.,  2015, \mn@doi [Data Science Journal]
  {10.5334/dsj-2015-011}, 14, 11

\bibitem[\protect\citeauthoryear{Zhu, Dai, Bian, Chen, Chen  \& Hu}{Zhu
  et~al.}{2019}]{zhuGalaxyMorphologyClassification2019}
Zhu X.-P.,  Dai J.-M.,  Bian C.-J.,  Chen Y.,  Chen S.,   Hu C.,  2019, \mn@doi
  [Astrophys Space Sci] {10.1007/s10509-019-3540-1}, 364, 55

\makeatother
\end{thebibliography}







\end{document}